\PassOptionsToPackage{table,xcdraw,svgnames,dvipsnames}{xcolor}
\documentclass[9pt,sigconf,letterpaper]{acmart}

\AtBeginDocument{\providecommand\BibTeX{{\normalfont B\kern-0.5em{\scshape i\kern-0.25em b}\kern-0.8em\TeX}}}

\usepackage{balance}

\newcommand{\ie}{i.e., \@}
\newcommand{\eg}{e.g., \@}

\newcommand{\nmap}{Nmap\xspace}

\newcommand{\eat}[1]{}

\newcommand{\curlie}[1]{Curlie}

\newcommand{\tool}{LFP\xspace}
\newcommand{\eid}{engine ID\xspace}
\newcommand{\Eid}{\expandafter\MakeUppercase \eid}

\newcommand{\Eids}{{\expandafter\MakeUppercase \eid}s\xspace}
\newcommand{\etime}{engine time\xspace}
\newcommand{\Etime}{\expandafter\MakeUppercase \etime}
\newcommand{\eboots}{engine boots\xspace}
\newcommand{\Eboots}{\expandafter\MakeUppercase \eboots}
\newcommand{\lastreboot}{last reboot time\xspace}
\newcommand{\Lastreboot}{\expandafter\MakeUppercase \lastreboot}

\usepackage[normalem]{ulem}
\usepackage{savesym}
\savesymbol{Bbbk}
\usepackage{tabularx}
\newcolumntype{L}[1]{>{\raggedright\arraybackslash}p{#1}}
\newcolumntype{C}[1]{>{\centering\arraybackslash}p{#1}}
\newcolumntype{R}[1]{>{\raggedleft\arraybackslash}p{#1}}
\usepackage{graphicx}
\usepackage{subfig}
\usepackage{float}
\usepackage{subfig}
\usepackage{enumitem}
\setlist{nolistsep}
\usepackage{graphicx}
\usepackage{epstopdf}
\usepackage{grffile}
\usepackage{amssymb}
\usepackage{amsmath}
\usepackage{rotating}
\usepackage{url}
\usepackage{enumitem}
\usepackage{color}
\usepackage{xspace}
\usepackage{ifpdf}
\usepackage{mdwlist}
\usepackage{colortbl}
\usepackage{caption}
\usepackage{amssymb}
\usepackage{booktabs}
\usepackage{multicol}
\usepackage{multirow}
\usepackage{enumitem}
\usepackage{xfrac}
\usepackage{afterpage}\usepackage{pdflscape}
\usepackage{longtable}
\usepackage{hhline}
\usepackage{lscape}
\usepackage{pifont}
\usepackage{fancyvrb}

\usepackage{arydshln} 

\usepackage[absolute,showboxes]{textpos} 

\usepackage{hyperref}

\usepackage{cleveref}

\graphicspath{{figures/}}

\copyrightyear{2023}
\acmYear{2023}
\setcopyright{rightsretained}
\acmConference[IMC '23]{Proceedings of the 2023 ACM Internet Measurement 
Conference}{October 24--26, 2023}{Montreal, QC, Canada}
\acmBooktitle{Proceedings of the 2023 ACM Internet Measurement Conference 
(IMC '23), October 24--26, 2023, Montreal, QC, Canada}
\acmDOI{10.1145/3618257.3624813}
\acmISBN{979-8-4007-0382-9/23/10}

\makeatletter
\gdef\@copyrightpermission{
	\begin{minipage}{0.3\columnwidth}
		\href{https://creativecommons.org/licenses/by/4.0/}{\includegraphics[width=0.90\textwidth]{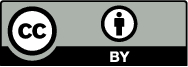}}
	\end{minipage}\hfill
	\begin{minipage}{0.7\columnwidth}
		\href{https://creativecommons.org/licenses/by/4.0/}{This work is 
		licensed under a Creative Commons Attribution International 4.0 
		License.}
	\end{minipage}
	\vspace{5pt}
}
\makeatother

\begin{document}

\title{Illuminating Router Vendor Diversity Within Providers and Along Network Paths}

\author{Taha Albakour}
\affiliation{\institution{TU Berlin}
\country{}
}
\author{Oliver Gasser}
\affiliation{\institution{Max Planck Institute for Informatics}
\country{}
}
\author{Robert Beverly}
\affiliation{\institution{Center for Measurement and Analysis of Network Data}
\country{}
}
\author{Georgios Smaragdakis}
\affiliation{\institution{Delft University of Technology}
\country{}
}

\renewcommand{\shortauthors}{Taha Albakour, Oliver Gasser, Robert Beverly, \& 
Georgios Smaragdakis}

\settopmatter{printacmref=true, printccs=true, printfolios=true}
\pagenumbering{gobble}

\begin{abstract}
\def\newabstract{yes}

\ifdefined\newabstract
The Internet architecture has facilitated a
multi-party, distributed, and heterogeneous physical infrastructure
where routers from different vendors connect and inter-operate via IP.
Such vendor heterogeneity can have
important security and policy implications. For example, a security
vulnerability may be specific to a particular vendor and
implementation, and thus will have a disproportionate impact on
particular networks and paths if exploited.  From a policy
perspective, governments are now explicitly banning particular
vendors---or have threatened to do so.

Despite these critical issues, the composition of router vendors 
across the Internet remains largely opaque.  Remotely identifying
router vendors is challenging due to their strict security
posture, indistinguishability due to code sharing across vendors,
and noise due to vendor mergers.  
We make progress in overcoming
these challenges by developing \tool, a tool that improves the
coverage, accuracy, and efficiency of router fingerprinting as compared to the
current state-of-the-art. 
We leverage \tool to characterize the degree of router
vendor homogeneity within networks and the regional distribution of
vendors.  We then take a path-centric view and apply \tool to better
understand the potential for correlated
failures and fate-sharing.  Finally, we
perform a case study on inter- and intra-United States data paths to
explore the feasibility to make vendor-based routing policy decisions,
\ie whether it is possible to avoid a particular vendor given the
current infrastructure.  
\else 
Remote router fingerprinting has many applications ranging from the assessment
of potential impact due to vendor vulnerabilities to the estimation of vendors'
market share to a better understanding of router deployment strategies at
different parts of networks or regions around the globe. Previous studies showed
that analyzing the response to a single-packet unsolicited and unauthorized
SNMPv3 request made it possible to collect vendor-level labels for around 20\%
of the routers reliably.

In this paper, we show that by sending nine additional probe packets per router
IP, it is possible to double the coverage of the routers that can be
fingerprinted in the wild, reaching around up to 40\% to 60\%. Our proposed technique is
more stealth, scalable, and accurate compared to other popular network scanners
like \nmap that require sending orders of magnitude more probe packets.  Our
analysis shows that with our proposed classification, it is possible to label
routers in regions and networks that were not possible with previously proposed
techniques. Moreover, in more than 50\% of paths we studied, we can infer the
vendors of routers along the path and thus, inform routing decisions.

\fi
 \end{abstract}



\maketitle

\section{Introduction}\label{sec:introduction} 

The Internet is exemplified by distributed control, varied policies,
and autonomy. This inherent heterogeneity extends to the physical 
infrastructure.  We explore one specific
physical property in detail: the composition of Internet router
\emph{vendors}.  The set of vendors through which data packets
traverse end-to-end has both direct and nuanced implications, for
instance reliability and fate-sharing, but also security and policy when a
vendor is untrusted.  Indeed governments have explicitly forbidden
particular vendors, while others have threatened to do so
\cite{Huawei-is-PRC-spy-ops}.

Unfortunately, the distribution and composition of router vendors
across the Internet remains largely opaque.  Operators consider their
network configurations sensitive and proprietary, and do not publicly
publish information on vendors.  While remote vendor and operating
systems fingerprinting is common for end hosts, \eg \nmap~\cite{Nmap},
these techniques rely on active services and responsive network
stacks.  However, routers are typically configured with strict
security to block arbitrary
requests~\cite{BackDoorIPv6,IPv4Allports}.  Indeed, multiple security
incidents have demonstrated the value in router fingerprinting and
reconnaissance in mounting
attacks~\cite{mirai-botnet-usenix-security,ZMap,Compromized-Routers}. 

While general purpose remote fingerprinting tools have been used to
great effect, \eg for Internet-wide surveys and
scanning~\cite{Ps-Qs,TTL-fingerprinting,TCP-revisited}, these tools
are ill-suited to router fingerprinting. 
Existing techniques are frequently unable to make a reliable vendor 
inference and 
typically send a
large number of packets making 
them (i)
impractical to operate at a large scale and (ii) intrusive. 

We leverage a
technique based on SNMPv3 which by itself is highly accurate, but provides poor
coverage.  Our primary insight is to collect lightweight network and transport
layer fingerprints and use this prior technique as a 
source of ground-truth vendor labels---this allows us to find new 
feature sets that uniquely identify vendors even when they do not
respond to SNMPv3.  In this fashion we more than double the
coverage compared to the SNMPv3
technique.
By tuning our technique to routers, we achieve this coverage while simultaneously 
improving efficiency by sending approximately two orders of magnitude
fewer
active probe packets 
as compared to existing
state-of-the-art that sends up to thousands of packets per inference. Our 
contributions include:

\begin{itemize}[leftmargin=*]
\item  {\it Lightweight FingerPrinting (\tool)}, a novel lightweight, 
accurate, and
more complete  remote vendor
fingerprinting methodology and tool
(\S\ref{sec:methodology}).

\item Using SNMPv3 to find sets of TCP/IP stack features for
\tool
that uniquely identify 
82\% of ground-truth routers and provide 95\% accuracy alone in 
fingerprinting 
major router vendors (\S\ref{sec:active}).

\item Applying \tool to the widely used 
RIPE
Atlas and CAIDA Internet Topology Data Kit (ITDK) router datasets
to classify 
64\% of the active router
IPs---more than double the coverage as compared to current state-of-the-art
(\S\ref{sec:analysis}).

\item An accuracy evaluation of \tool compared to current tools and techniques
showing 
that it is at least as good as \nmap 
while sending orders of magnitude fewer packets and improving coverage
(\S\ref{sec:analysis:comparison}).

\item Inference of router vendors in more
than 6,700 
networks, including around 
1,800 
networks for which no vendor  information 
was 
available in previous studies
(\S\ref{sec:analysis:deployment}). 

\item End-to-end data path-based router vendor analysis and case studies
that provide valuable insights for current security and policy-based
routing decisions (\S\ref{sec:paths}).

\item \tool is publicly available, along with the 
the derived signatures and classification results from this
study to enable reproducibility and future work ~\cite{IMC2023-artifacts}. 

\end{itemize}

 \section{Related Work} 

Prior work developed passive and active techniques
that leverage open ports, identifiers, and implementation-specific differences to
fingerprint devices at various granularities. Most of these techniques were
developed for generic hosts, while a few attempt to fingerprint routers.

\noindent\textbf{Nmap:} Nmap \cite{Nmap} is an open-source network scanning and
reconnaissance tool. It performs remote OS fingerprinting
by running up to sixteen tests that send
ICMP, UDP, and TCP packets with different field
values, flags, and options.  By examining the
responses, Nmap finds the best matching operating system from a  
database of fingerprints. The latest Nmap version (7.93) 
contains more than six thousand fingerprints; of these, approximately 160 and 20 correspond to
Cisco and Juniper routers, respectively.
Two drawbacks of Nmap are its reliance on open ports and the large amount of probe packets needed to perform fingerprinting.
We compare our approach against Nmap in
(\S\ref{sec:analysis:comparison:nmap}).

\noindent\textbf{Hershel:} Hershel~\cite{Hershel} is a low-overhead framework
that models the problem of single-packet OS fingerprinting and develops novel
approaches for tackling delay jitter, packet loss, and user modification to
SYN-ACK features. Based on this theory, a classification method is developed that
increases the accuracy of single-packet fingerprinting. Censys~\cite{Censys} includes Hershel
signatures in recent raw scanning data.
We compare \tool to Hershel in (\S\ref{sec:analysis:comparison:hershel}).

\noindent\textbf{Banner Grabbing:} A popular technique for remote operating system
fingerprinting and vendor information is ``banner grabbing,'' whereby publicly
available services leak information. For instance, the Cisco SSH server
implementation returns identifying information in its response string.
Internet-wide scanning and banner grabbing are performed
regularly~\cite{ZMap,LZR2021,Censys,IPv4Allports}.
In a recent paper~\cite{Classifying-Vendors}, the authors utilize banners
augmented with active measurements to perform
large-scale network equipment vendor classification. Similar techniques can be
applied in passive measurements as well for automatic traffic
classification~\cite{new-directions-ccs21}. Unfortunately,
banner analysis requires an open remote service that
returns this discriminating information. Routers are frequently tightly secured and
unresponsive to banner queries. Moreover, banner datasets are typically
proprietary or commercial, with some offering free academic
licenses~\cite{Censys}. 

\noindent {\bf TCP Stack Fingerprinting:} Many TCP stack variables, e.g., Window
Size and Maximum Segment Size are
implementation-specific~\cite{beverly-classifier,tcp_rfc,IETF-RFC4413}.  These
variables can differ between operating systems and versions. Consequently, TCP features can form a unique signature that can be used
for fingerprinting.  For instance, the initial TCP
SYN-ACK packet provides some valuable information about a target's TCP stack
characteristics such as the initial Time to Live (TTL) value, sequence number,
and window size.  When combined with the behavior of the re-transmission timeout
of the SYN-ACK packets it was shown to serve as a fingerprinting technique to 25
different operating systems~\cite{noauthor_snacktime_nodate}, and in another work
this was extended to cover more than 90 OSes~\cite{leonard_demystifying_2010}.

\begin{figure*}[!tpb]
    \centering
	\includegraphics[width=.9\linewidth]{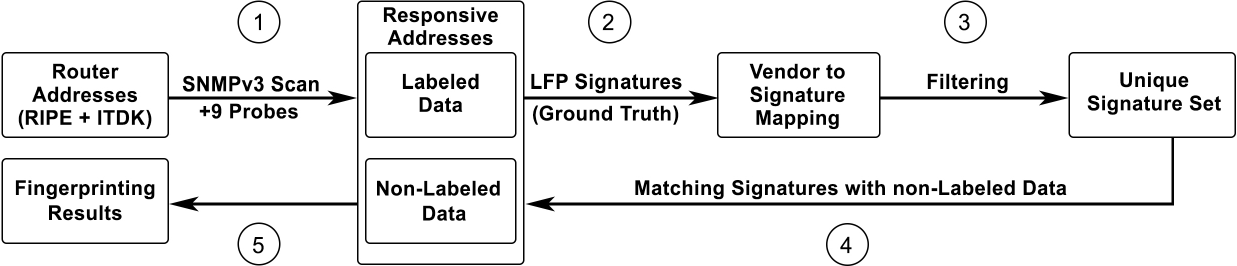}
	\caption{Data Collection Pipeline.}
	\label{fig:pipeline}
\end{figure*}

\noindent {\bf Sundials:} Sundials~\cite{Sundials-PAM2019} 
uses ICMP timestamps for fingerprinting purposes.
Even though NTP has replaced ICMP timestamps, 
approximately 15\% of 14.5M IP addresses in this study responded to ICMP timestamp requests.
Sundials uses the variety of response behaviors as a new fingerprinting technique.
However, given filtering and the relative lack of ICMP timestamp
support among routers, this method has
limited coverage for our fingerprinting purposes.

\noindent\textbf{IPID-based Fingerprinting:} The IP identifier (IPID)
is a mandatory IPv4 header field used for fragmentation and
reassembly. Thus, it is frequently possible
to elicit an IPID value from a router via a simple ICMP echo.
RFC 4413~\cite{IETF-RFC4413} classifies IPID behavior into three classes:
{\it (1) Sequential jump}: an incremental IPID counter that is used for all
packet streams, or
{\it (2) Random}: a pseudo-random number generator is used for the IPID value, or
{\it (3) Sequential}: an incremental IPID counter on a per-stream basis. 
The IPID may also have a static value, \eg zero.
While the limited size (16 bits) of the IPID counter can be
problematic,
Internet researchers have utilized the IPID field for a broad range of
applications. Bellovin~\cite{bellovin_ipid_nat} uses IPID to count NATed hosts, alias resolution
tools such as MIDAR~\cite{keys13midar} and Ally~\cite{Rocketfuel} use monotonic IPID counters
to infer aliases, and Chen et al.~\cite{chen_exploiting_2005} use
IPIDs to characterize end-systems.
In this work, we utilize the differences in IPID value generation between
router vendors across protocols for fingerprinting purposes. 

\noindent\textbf{TTL-based Fingerprinting:} Vanaubel et al.~ propose a router fingerprinting technique
based solely on TTL responses~\cite{TTL-fingerprinting}.
They send TCP, UDP, and ICMP probes toward the target, and show that the tuple
of inferred initial TTL (iTTL) values from the responses can coarsely differentiate
between some well-known vendors, including Cisco and Juniper.  Unfortunately,
the possible iTTL value range is small, and can lead to a large number of
incorrect inferences, \eg we find that Huawei has the same iTTL signature as Cisco.
Nevertheless, we use the iTTL values as part of a larger feature set.

\noindent {\bf SNMPv3-based Fingerprinting:} Most recently, research
has shown that  
the adoption of the SNMPv3
protocol offers an opportunity for remote fingerprinting of
network infrastructure~\cite{IMC2021-SNMPv3} including routers. In addition to gathering detailed information about network devices, such as
vendor, uptime, and the number of restarts, the reply also contains a
strong, persistent identifier that allows for lightweight alias resolution and
dual-stack association. We leverage this SNMPv3 technique to build a ground 
truth, and use it as a baseline for our proposed \tool method.

 \section{Methodology}\label{sec:methodology}

This section presents our methodology to scalably classify routers in the wild at
vendor granularity. We first give an overview,
then describe our dataset, measurement probes, and the features we use for 
classification.
Subsequently, we generate signatures based on these features and we describe how we handle classification edge cases due to ambiguity or lack of data.
Finally, we elaborate on the limitations of our methodology.
We refer the reader to \Cref{sec:ethics} for the ethical principles guiding our measurements.
For a pipeline of our methodology, we refer the reader to Figure~\ref{fig:pipeline}.

\subsection{Overview}

Our methodology builds a model based in part on high-confidence router vendor
labels and then uses that model to extend coverage and improve
accuracy.  Such ground-truth
data can be collected using private information about the deployment of 
routers
in a network or via information leakage using protocol scanning~\cite{Censys}.
In our method, we utilize SNMPv3 as described in 
~\cite{IMC2021-SNMPv3} which is able to accurately label around 20\% of
routers in the wild by sending a single unsolicited and unauthenticated
request. The reply to this
request contains detailed information including a router ``Engine
ID.''  This ID
easily and reliably identifies the router vendor.  We leverage
the SNMPv3 technique and scan a set of router addresses to create labeled 
data and build a classification model for router fingerprinting measurement.

We expect that, typically, routers will not expose services to the public
Internet. We decide to use three types of probe packets over ICMP, TCP,
and UDP. ICMP has been used before to
fingerprint routers, as many routers respond to ICMP packets. For TCP
and UDP, we expect that routers do not expose such services to the public
Internet. However, the response to packets targeting a closed port for these
protocols can provide useful information towards fingerprinting the router
vendor. In addition to the SNMPv3 requests, 
we send three single-packet probes over each of the
three primary transport protocols, namely, ICMP, TCP, and UDP for a 
total of nine probes per IP address (Figure~\ref{fig:pipeline} 
\textcircled{1}). We explain the rationale
to use these three protocols in Section~\ref{sec:features}.  
The feature values of the
responses to our transport protocol measurements, listed in 
\Cref{table:features}, are used to build a signature database for {\it 
Lightweight Fingerprinting (LFP)}. 
For IPs that are responsive 
to SNMPv3 requests, we extract the vendor information and used it as a label 
(Figure~\ref{fig:pipeline} \textcircled{2}). 
Note that our methodology is not dependent on the SNMPv3 to label routers. In
principle, any reliable router label source can be used as input to our
classification method.

\subsection{Datasets}

To select target router IPs, we leverage two complementary
public router
datasets: the RIPE Atlas traceroutes dataset \cite{RIPE-Atlas} and the ITDK 
dataset \cite{itdk}. We list the router datasets with dates, address
counts, and AS
coverage in \Cref{table:datasets}.

\noindent{\bf RIPE Atlas Traceroutes.}
We extract intermediate IP hops from RIPE Atlas traceroute measurements to obtain 
router IPv4 addresses.  We explicitly ignore the last responsive hop, if it is the
same as the targeted host IP, to ensure that we only include router IPs.  We
utilize five snapshots of traceroute data over a ten-month period from 
January -- November 2022. We extract between 446k to 496k router IPs from 
each snapshot. Further, each snapshot covers between
18.3k to 20.2k ASes. 
In order to increase the coverage, we utilize
all five snapshots to gather signatures and evaluate their stability over time.
Moreover, we find that RIPE Atlas traceroutes are relatively stable
across the ten-month period, with a pairwise router IP overlap of about 88\%
between consecutive collections.
Therefore, we utilize the most recent RIPE Atlas snapshot, \ie RIPE-5, for 
our IP level analysis.

\noindent{\bf ITDK Router-Level Topologies Dataset.}
In addition to IP level traceroute data, we also use the 
router topology from CAIDA's 
February 2022 ITDK~\cite{itdk}. 
This complementary dataset contains router alias sets (excluding singletons) 
inferred via MIDAR \cite{keys13midar} and iffinder measurements.
This dataset covers fewer IP addresses and about half of the number of ASes 
compared to RIPE Atlas. This is expected as addresses in this dataset must 
respond over at least one protocol (ICMP, UDP, or TCP) which is required 
for alias resolution. This is also evident in our active measurement where we 
note a higher responsiveness for the ITDK data compared to RIPE Atlas as 
shown in Figure~\ref{fig:responsiveprotos}.
The complementary nature of this dataset is underscored by a relatively
low overlap of at most 26\% router IPs present in any of the RIPE Atlas 
traceroute 
datasets. 
We use the ITDK data for gathering signatures and router level analysis.

The union of all RIPE Atlas traceroute and ITDK MIDAR datasets covers more than
970k router IP addresses in about 25k ASes. We note that our methodology is not
limited to these selected datasets, but in fact other datasets containing
candidate router IP addresses could be used as well.  Next, we run active
measurements toward targets in each of these datasets to gather
features and build signatures
for router fingerprinting.

\begin{table}[!bpt]
	\caption{List of features used with possible values.}
        \vspace{-2mm}
	\label{table:features}  
	{
		\resizebox{\columnwidth}{!}{
			\begin{tabular}{ll}
                \toprule
				Feature & Possible Values \\
                \midrule
				ICMP IPID echo & true, false \\ 
				ICMP IPID counter & incremental, random, static, zero, duplicate \\ 
				TCP IPID counter & incremental, random, static, zero, duplicate \\ 
				UDP IPID counter & incremental, random, static, zero, duplicate \\ 
				TCP UDP ICMP shared counter & true, false \\ 
				TCP ICMP shared counter & true, false \\ 
				UDP ICMP shared counter & true, false \\ 
				TCP UDP shared counter & true, false \\ 
				UDP iTTL & 32, 64, 128, 255 \\ 
				ICMP iTTL & 32, 64, 128, 255 \\ 
				TCP iTTL & 32, 64, 128, 255 \\ 
				ICMP echo response size & variable \\ 
				TCP response size & variable \\ 
				UDP response size & variable \\ 
				TCP SYN sequence number & zero, non-zero  \\
				\bottomrule
	\end{tabular}}}
\end{table}

\begin{table}[!bpt]
	\caption{Overview of router address datasets: Number of unique IP 
	addresses and Autonomous Systems. We utilize all data sources for 
	signatures gathering. However, we use RIPE-5 for path and IP-level 
	analysis 
	and ITDK for router analysis.}
	\label{table:datasets}  
	{\small
		\resizebox{\columnwidth}{!}{
			\begin{tabular}{llrr}
				\toprule
				Data Source & Date & \# IPv4 addrs. &  \# ASes\\
				\midrule                
				RIPE-1 & 2022-01-24 & 494,867 & 20,178 \\
				RIPE-2  & 2022-02-24 & 484,930 & 19,989 \\
				RIPE-3 & 2022-06-09 & 496,167 & 20,085 \\
				RIPE-4 & 2022-07-04 & 446,629 & 18,304 \\
				RIPE-5 & 2022-11-07 & 476,577 & 18,837 \\ 
				ITDK  & 2022-02 & 343,312 & 9,922 \\
				\midrule
				Union & & 971,343 & 24,909 \\ 
				\bottomrule
	\end{tabular}}}
\end{table}

\subsection{Active Scanning Packets}

To collect router fingerprints, we send 10 packets in total per target
IP: 3
for each transport protocol and a single SNMPv3 request. 
We aim to reduce the impact of our scan on the target by using simple ping probes and avoiding any malformed packets.
For ICMP, we send three echo
requests. For each of these requests we expect an echo reply. For TCP and UDP we
target port 33533, with the assumption that no services are active on this port.
For TCP we send two ACK packets and one SYN packet with a non-zero acknowledgment number. 
We expect that all three TCP packets - both the ACK and SYN - will elicit a TCP RST response.
For UDP we send three packets, each with 12 bytes of all zero payload. For each packet, we expect to receive an ICMP port
unreachable response. 

\subsection{Feature Set}\label{sec:features}

We limit our methodology to features that can be extracted mainly from the IP layer.
In total, we extract 15 features from our 9 probe packets (see Table~\ref{table:features}).
We consider four groups of features:

\subsubsection{IPID} 

We send a trio of consecutive packets and collect the IPID values from all 
responses. We then construct IPID sequences for each protocol.
Previous work~\cite{keys13midar,Rocketfuel}
showed that IPID sequences exhibit distinct patterns, e.g., they can be monotonically increasing
or random. These patterns can not only be used to perform IP-alias resolution as shown in previous work, but
also facilitate the identification of a router's vendor. One test for device fingerprinting is checking if ICMP request and response IPID values match~\cite{Nmap,arkin2002xprobe}. IPID sequences differ among different protocols, but some implementations use the same sequence across all protocols. As we show in Table~\ref{table:features}, the ICMP IPID Echo feature indicates whether Echo request and response IPID values are the same (true) or different (false).
The IPID counter for any of the three protocols (ICMP, TCP, and UDP) can be
characterized as incremental (which can also include wrap-around from the
largest 16 bit value back to starting at zero), random, static (always the same
value, other than zero), zero (always responds with an IPID of zero), or with
duplicates (where exactly two responses have the same IPID value).

\subsubsection{iTTL}

Previous work~\cite{TTL-fingerprinting} showed that different initial TTL (iTTL)
values may differ between different protocols and even message types. We collect
the iTTL values for each response that we receive.
Typically, the iTTL value depends on the operating system or network card per
vendor. Indeed, in \Cref{table:features}, we show the different values,
four in total, that we have collected in all our experiments (see
Section~\ref{sec:active} for details). 

\subsubsection{Response Size}

To further diversify our features, we collect the response size for all
protocols. We notice that typically, the ICMP and TCP response size often do not
provide any information gain. However, the ICMP port unreachable response to a
UDP request packet can differ between router vendors. This depends on whether
the request packet is fully or partially quoted (and if so, how much of the
original packet is quoted) in the ICMP response packet \cite{rfc792,rfc1812}.
As we show in~\Cref{table:features}, the characteristic value is variable
and differs by router vendor and implementation, which allows us to make use of
the response size for router fingerprinting.

\subsubsection{Additional Features}

RFC 793~\cite{tcp_rfc} states that if a port is closed, any incoming segment
except a reset triggers a reset response. If the segment has an ACK field, the
reset takes its sequence number from the ACK field, otherwise, it uses a
sequence number of zero.  We noticed that only a few vendors are compliant with
the RFC in this regard.

For the set of features and the possible feature values we refer to
Table~\ref{table:features}. 
We note that most of these features are only available for IPv4. Thus, in this
paper we focus only on the classification of IPv4 router interfaces.

\subsection{Signatures}
\label{sec:methodology:sub:signatures}

We assemble all responses for each IP address into a feature vector based on \Cref{table:features}.
We use the instances of a particular feature vector that are associated with a 
vendor obtained from the SNMPv3 probes to create a mapping of a feature 
vector to a vendor. We then apply a basic filter based on occurrences 
threshold  as described in (\S\ref{sec:occthreshold})
At this point the feature vector is used as a {\it signature} for the vendor
(Figure~\ref{fig:pipeline} \textcircled{3}). 

\noindent{\bf Unique Signatures.}
If a signature is mapped only to a single vendor, then we call this a {\it unique
signature}. In this case we have high confidence in the accuracy of the
signature.

\noindent{\bf Non-Unique Signatures.}
When a signature is associated with {\it multiple} vendors, we characterize this
as a {\it non-unique signature}. This may happen, e.g., due to the change
of the default router configuration by network operators, or simply a shared
TCP/IP stack implementation between multiple vendors. 
As we will show in Section~\ref{sec:active}, typically there is
one vendor that dominates even for non-unique signatures, or the non-unique
signatures map to a family of routers that are based on the same OS or network stack. However, for the purpose of this study, we take a conservative approach and only
consider {\it unique} signatures in our analysis.

\noindent{\bf Partial Signatures.}
There are also cases where a router IP responds only to a subset
of the all three protocols (ICMP, TCP, and UDP). In this case, we characterize
the signature as a {\it partial signature}. Even partial signature may prove useful to
identify the vendor of a router. If the partial signature is unique for a
vendor, then we call this a {\it partial unique signature}. If this partial
signature is associated with multiple vendors, we call it a {\it partial
non-unique signature}.

As we elaborate in detail in Section~\ref{sec:active}, it is common
for a single vendor
to have multiple signatures. This is to be expected as vendors often offer multiple
products and versions of the same product, or it can be an artifact of acquisitions. 
Once the list of signatures for a given vendor
has been compiled, we can match the signatures using our active measurements to
infer the vendor of unlabeled routers (Figure~\ref{fig:pipeline} \textcircled{4}).
With this technique we can substantially increase the coverage of routers that we can fingerprint
in the wild (Figure~\ref{fig:pipeline} \textcircled{5}).

\subsection{IPID Threshold}\label{sec:ipidthreshold}

\begin{figure}[!bpt]
	\centering
	\includegraphics[width=.7\linewidth]{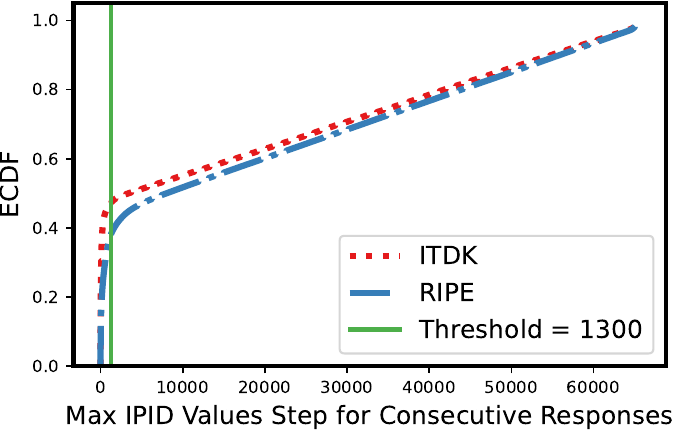}
	\caption{Maximum IPID step distribution per IP address. The vertical line 
	shows the chosen threshold between sequential and random IPID increases.}
	\label{fig:ipid_step}
\end{figure}

In order to determine whether an IPID counter is incremental or random, we 
investigate
the returned IPID values per IP address and across all three protocols. 
Consequently, we sample the IPID values only for fully responsive addresses. 
We calculate the step values for each 
consecutive packet pairs and aggregate them by applying a maximum
function\footnote{We obtain similar results when applying an average function
instead of a maximum function. Since the maximum function is more conservative,
we use it in our methodology.}. In \Cref{fig:ipid_step} we show the 
distribution 
of maximum IPID step per IP in the responses to all three protocols. 
In order to distinguish random from
sequential increases, we check for a knee in the distributions of
\Cref{fig:ipid_step}.  We empirically take a conservative threshold value of
1,300 to distinguish between sequential and random IPID counters. 
Note that a sequential increase can be larger
than `1', as concurrent traffic from that router also leads
to an increase in sequential IPIDs.  

We evaluate the empirical threshold by estimating the probability
of misclassifying a random IPID counters as a sequential. Recall that we sent 
9 packets in total and calculated 8 IPID steps by determining the difference 
between two consecutive IPID values. Given our threshold, the probability of 
a random IPID counter generating a value less than or equal to the threshold 
is {1301}/{$2^{16}$} which is $\approx$ 0.019. For our classifier to 
misclassify a random counter as sequential, all eight IPID steps need to be 
less than or equal to the threshold, which has an extremely low probability 
of $0.019^{8}$ when considering all protocols, or $0.019^{2}$ when 
considering each protocol separately.

We further explore the empirical threshold in~\Cref{fig:IPID-diff-slide19}, 
where we plot the distribution of the IPID difference
for consecutive responses for fully responsive IPs in the RIPE-5 dataset. It
is clear that around 20\% of IPID differences are close to zero.
Close to 90\% of the IPID difference values are included
by setting a threshold of 1,300, as shown with the dashed vertical line
We use this threshold to differentiate between incremental values and
random that are dispersed across the full range of possible IPID difference
values.

\begin{figure}[!tbp]
	\centering
	\includegraphics[width=.7\linewidth]{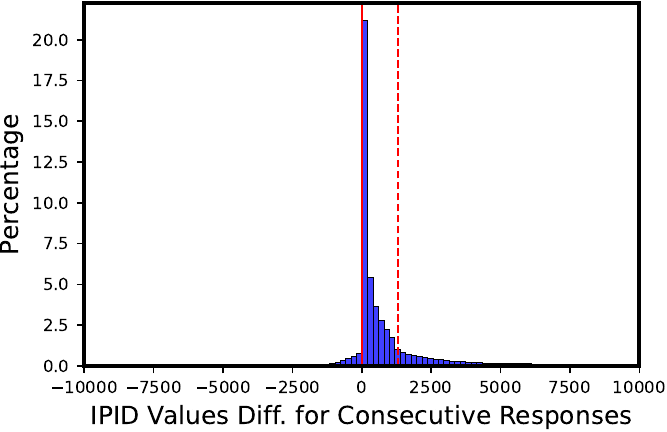}
	\caption{Distribution of IPID difference values for consecutive 
	responses.}
	\label{fig:IPID-diff-slide19}
\end{figure}

Note that an effectively random IPID might by chance fall within this 1,300 threshold.
Since with LFP we take the conservative approach of using the maximum IPID difference between consecutive probes, this random effect is very unlikely to occur twice in a row and thus strongly minimizes the number of false positives.

Finally, we also characterize the sharing of IPID sequences across pairs or all
protocols as true of false, if this takes place or not, respectively.

\subsection{Limitations}

\tool improves the state-of-the-art in remote router vendor
fingerprinting.  
However, we acknowledge that several limitations remain: 
\begin{itemize}[leftmargin=*]
\item Our classification builds on highly accurate vendor data
obtained via SNMPv3 probing, however SNMPv3 coverage is not
universal
and imparts bias.  While
the SNMPv3 technique obtains correct labels 
for approximately 20\% of
the routers and 30\% of the router IPs we probed, 
we do not generate a signature for vendors that do not implement or
do not respond to
SNMPv3 requests.
This results 
in a bias toward SNMPv3 enabled routers and can negatively affect the 
uniqueness of generated signatures. 

\item As we elaborate in Section~\ref{sec:active}, a non-negligible
fraction of routers do not respond to \emph{any} remote probe. This differs across 
sets of
router datasets, but for these routers our technique cannot provide any
insights.

\item New signatures may be created as 
as new router models or
vendors are introduced in the market. Although in Section~\ref{sec:active}
we show that over a period of ten months, and for different router
datasets, the
signatures we discover remain stable, over longer period of time, e.g., years, 
new measurements may be required to keep \tool signatures up-to-date.

\item We restrict our analysis to core routers. 
A primary challenge to
fingerprinting edge routers is the greater 
diversity of Customer Premise Equipment (CPE) and residential
equipment, along with substantial amounts of IP churn.
Although we believe that our
technique can be used to fingerprint edge network equipment, we
defer such an investigation to future work. 

\item We may misclassify random IPID response sequences as sequential.
To significantly minimize the potential for erroneous inference, we 
we take the maximum IPID step difference among all pairwise IPID values
(see \Cref{sec:ipidthreshold}).

\item We focus our study on the IPv4 Internet. Many of the features that
\tool relies on (see \Cref{table:features}) are not available in IPv6 or
do not provide the same discriminatory opportunities for fingerprinting.
For instance, the IPv6 header does not include an IPID field unless 
fragmentation is induced \cite{Speedtrap}, rendering all IPID-related 
features inapplicable 
for fingerprinting. Furthermore, all IPv6 implementations use the recommended 
initial TTL value of 64 \cite{IPv4-IPv6-Fingerprinting-ICMP}. The remaining 
features do not provide significant 
information gain to produce an accurate vendor signature.

\item We limit the scope of our work and focus only on the technical aspects 
of 
remote router vendor fingerprinting
that can be used to inform routing decisions. We recognize that better insight
into vendors within ASes and along end-to-end paths is especially interesting
given the current climate where, e.g., some countries are imposing restrictions
on the use of equipment from particular vendors. In this paper, we discuss this
issue in a purely factual, impartial, and non-political manner.  Since we are
not legal or political science scholars, we do not discuss, opine, or speculate
on non-technical matters, e.g., the legal, financial, and social
impact of our work. 

\end{itemize}
 \section{Active Experiments}\label{sec:active}

\begin{table*}[t]
    \center
	\caption{Measurement overview: Responsive IPs (IPs), responsive IPs to SNMPv3 (SNMPv3), to SNMPv3 and LFP (SNMPv3 $\cap$ LFP), only to LFP (LFP $\setminus$ SNMPv3), number of unique signatures (Unique sigs), and non-unique signatures (Non-unique sigs).}
	\label{table:ipid-tests-overview}  
	{\begin{tabular}{lrrrrrrr}
			\toprule
			Measurement & IPs & SNMPv3 & SNMPv3 $\cap$ LFP & LFP $\setminus$ SNMPv3 & Unique sigs   & Non-unique sigs \\
			\midrule
			RIPE-1 & 359,263           & 99,560    & 55,116
& 58,266               & 62           & 9\\ RIPE-2 & 355,709          & 95,600    & 54,933            & 59,400
& 46           & 8 \\ RIPE-3 & 363,464          & 94,699    & 53,196             & 58,843
& 47           & 10 \\ RIPE-4 & 323,141           & 82,047    & 48,360             & 72,969
& 49          & 11 \\ RIPE-5 & 327,534           & 90,540    & 47,700             & 77,298
& 51           & 13 \\ ITDK & 311,607           & 113,089   & 58,492              & 53,952               & 34           & 7 \\
			\midrule
            Union & 736,260 & 218,129 & 132,524 & 169,143 & 89 & 23 \\ \bottomrule
	\end{tabular}}
\end{table*}

We now apply our LFP methodology in active experiments to fingerprint routers in the wild.
We run six measurements, one for each data source (five RIPE Atlas traceroutes, and one ITDK's MIDAR dataset, cf. \Cref{table:ipid-tests-overview}).
We find the five RIPE Atlas based measurements to be relatively consistent.
Between 82k and 100k IPs are responsive to SNMPv3.
Of those around 50k respond to all LFP probes, \ie our labeled dataset which 
we use to extract vendor information.
Another 58k--77k respond only to LFP probing, \ie our dataset that we can fingerprint with the LFP technique without the IPs responding to SNMPv3.
The ITDK dataset provides more SNMPv3-responsive IPs, with a similar number of LFP responses compared to RIPE Atlas traceroutes.

\subsection{Responsiveness}

\begin{figure*}[!ht]
	\begin{minipage}[c]{.32\linewidth}
		\centering
		\includegraphics[width=\linewidth]{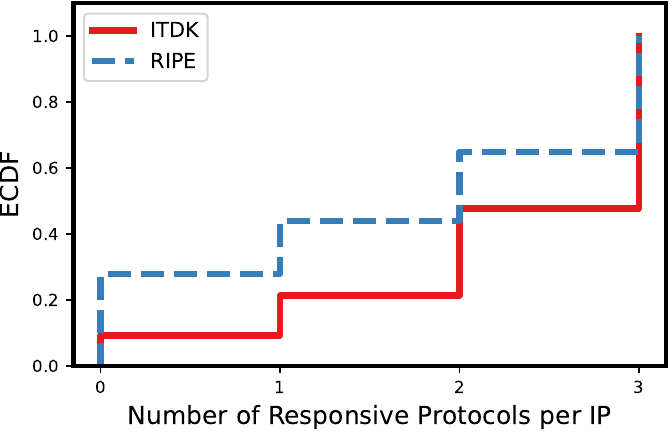}
		\caption{Responsive protocols per IP for the RIPE-5 and ITDK 
		datasets.}
		\label{fig:responsiveprotos}
	\end{minipage}
	\hfill \begin{minipage}[c]{.32\linewidth}
		\centering
		\includegraphics[width=\linewidth]{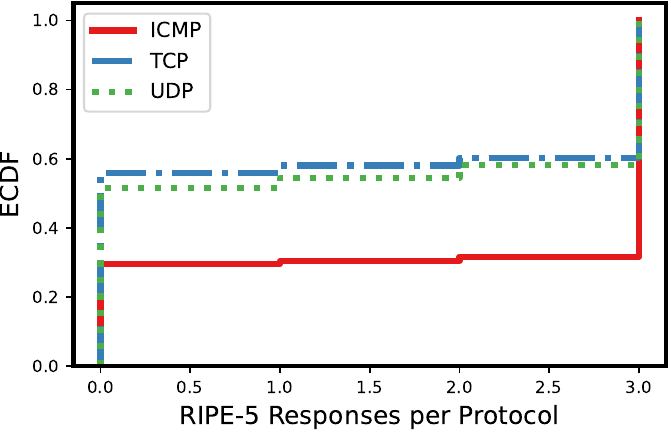}
		\caption{Responsiveness per protocol for the RIPE-5 dataset.}
		\label{fig:responsiveripe}
	\end{minipage}
	\hfill \begin{minipage}[c]{.32\linewidth}
		\centering
		\includegraphics[width=\linewidth]{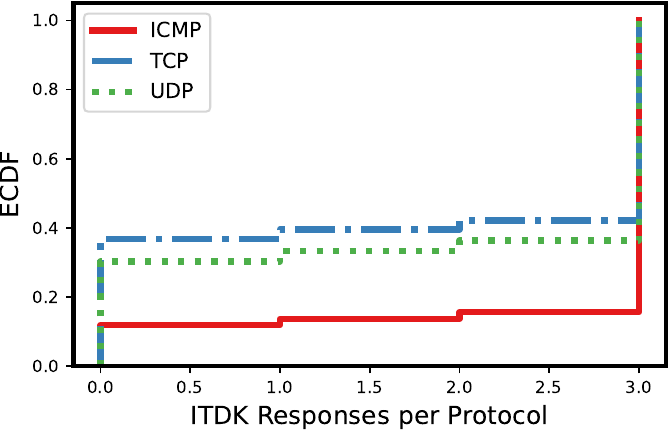}
		\caption{Responsiveness per protocol for the ITDK dataset.}
		\label{fig:responsiveitdk}
	\end{minipage}
\end{figure*}

Next, we analyze how responsive the target datasets are to LFP probes.
This determines the upper limit of our coverage with LFP.
\Cref{fig:responsiveprotos} shows the distribution of the number of responsive protocols (TCP, UDP, ICMP) per IP, comparing the ITDK and RIPE-5 datasets.
Since we rely on responsiveness to create signatures, the higher the number
of responsive protocols, the higher the entropy in the signature. 
Generally, we find that ITDK provides more responsive protocols compared to RIPE.
About 50\% of ITDK IPs are responsive on all three protocols, which is only 35\% for RIPE.
It is very encouraging, however, that we get responses from at least one protocol for 90.7\% and 72.3\% for ITDK and RIPE, respectively.

One other factor influencing the uniqueness of our signatures is the number of
responses \emph{per protocol}.  \Cref{fig:responsiveripe,fig:responsiveitdk}
show the responsiveness per protocol for RIPE-5 and ITDK, respectively.  In both
datasets we see that ICMP is more likely to elicit responses compared to TCP or
UDP.  Moreover, we see that an IP responds either to all three probe packets per
protocol or to none, \ie the line from zero to three in the plot is almost
horizontal.  Finally, we find that IP addresses from the ITDK dataset are
generally more likely to be responsive compared to the RIPE dataset: 84.4\% vs.
65.7\% are responsive on all three ICMP probes for ITDK and RIPE, respectively;
for TCP and UDP the difference is 63.6\% in ITDK compared to 39.5\% in RIPE.

\subsection{Signatures}

After collecting all responses from our measurements, we extract features (cf. \Cref{sec:features}) and create signatures based on our labeled SNMPv3 data (cf. \Cref{sec:methodology:sub:signatures}).
As can be seen in \Cref{table:ipid-tests-overview}, each dataset individually contributes 34--62 unique signatures and 7--13 non-unique signatures.
Unique signatures give us a high confidence when applying our LFP technique, as all labeled instances can be mapped to the same router vendor.
Note, that if the same unique signature would be found with different vendors in different datasets, we count it as a non-unique signature.
We find this case to be relatively rare, however, with only 2 occurrences for our five datasets.
In our fingerprinting analysis, we exclude any non-unique signature and use the
union of all five datasets to create a total of 89 unique signatures.
We set a threshold of a minimum of 20 router samples per signature.
Setting the threshold lower will only increase the covered routers by
1\%, but
disproportionally increases the number of signatures.

\subsection{Occurrences Threshold}\label{sec:occthreshold}

\begin{figure}[!tpb]
	\centering
	\includegraphics[width=.7\linewidth]{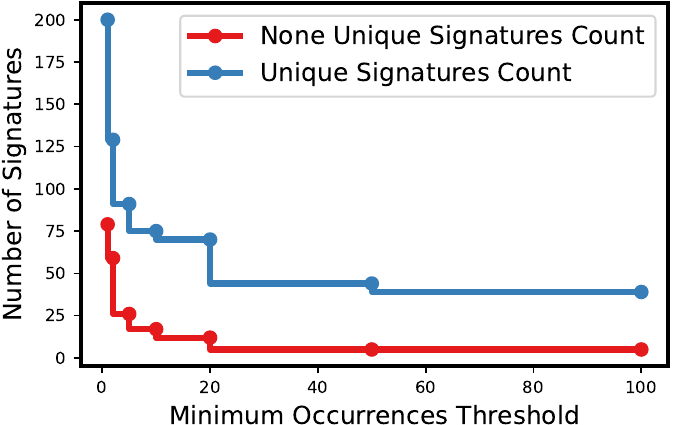}
	\caption{Sensitivity analysis: impact of setting a threshold on the number
		of occurrences for signatures on the number of unique and non-unique 
		signatures.}
	\label{fig:o-threshold-slide10}
\end{figure}

We perform a sensitivity analysis to understand the impact of the threshold of the
minimum number of occurrences for signatures to be considered.
In~\Cref{fig:o-threshold-slide10}, we vary the threshold (see x-axis),
i.e., the minimum number of IPs with the same signature. A low
threshold leads to a very high number of both unique and non-unique
signatures. This is to be expected as uncommon vendors or a small number of
configuration changes may lead to many different signatures.  However, when we
set the threshold of minimum occurrences to consider a signature to 10 or
more, the number of signatures, both unique and non-unique, converges. As
\Cref{fig:o-threshold-slide10} shows, choosing 10 or 20 as the threshold does not
change the number of signatures substantially. A closer investigation shows that
the set of signatures is also not affected.
Therefore, we choose a threshold of 20 router IPs per signature in our study as
it provides a good trade-off between considering a low number of popular
signatures and excluding a large number of rare signatures with only a few
occurrences in the hundreds of thousands of IPs in our dataset. In total, for
this study, we identified 89 unique signatures and 23 non-unique signatures.
We provide the full list of signatures in ~\cite{IMC2023-artifacts}.

\begin{table}[t]

	{\scriptsize
\center
	\caption{Partial signatures for different responsive protocol combinations.}
	\label{table:labeled-datasets-vendors-protocols}  
		\resizebox{.8\columnwidth}{!}{
			\begin{tabular}{lrrr}
				\toprule
				Protocols &  Total & Unique & Non-unique  \\
				\midrule
				TCP \& UDP & 61 & 43 & 18  \\
				ICMP \& UDP & 60 & 42 & 18  \\
				ICMP \& TCP & 51 & 36 & 15  \\
				UDP & 20 & 12 & 8  \\
				ICMP & 19 & 9 & 10 \\
				TCP & 17 & 10 & 7 \\
				\bottomrule
	\end{tabular}}}
\end{table}

In addition to signatures where we get responses from all protocols, we also
leverage partial signatures. \Cref{table:labeled-datasets-vendors-protocols} shows the partial fingerprints for different combinations of partial protocol responsiveness.
We find that if we see responses from two protocols (\ie TCP \& UDP,
ICMP \& UDP, or ICMP \& TCP), the majority of partial signatures are
still unique and can therefore be leveraged by the LFP technique.
Regarding single protocol signatures, the results are mixed. Most signatures are unique for TCP-UDP, ICMP-UDP and ICMP-TCP, while about half are unique for just TCP, UDP or ICMP. 
In general, utilizing unique partial signatures expands coverage
by $\approx$ 15\% while maintaining accuracy.

\subsection{Mapping Signatures to Vendors}

\begin{table}[t]
	\caption{Number of signatures in ground-truth dataset per router vendor.
(\#IPs are noted in parentheses).}
	\label{table:labeled-datasets-vendors}	
	{\small
		\resizebox{\columnwidth}{!}{
			\begin{tabular}{lrrr}
				\toprule
				Vendor & Labeled & Unique sigs & Non-unique sigs \\
				\midrule
				Cisco &  83,918 & 25 (82,020) & 1 (1,898)  \\
				MikroTik & 28,989 & 26 (9,489) & 4 (19,500) \\ 
				Huawei  & 19,869 & 8  (17,034)  & 4 (2,835)\\
				Juniper  & 17,665 & 15 (17,665)& 0 (0) \\
				H3C & 2,469 & 5 (358) & 5 (2,111) \\
				Alcatel/Nokia & 1,111 & 2 (1,111) &  0 (0) \\
				Ericsson & 200 & 1 (200) & 0 (0) \\
				Other & 9,676 & 4 (497) & 18 (9,179) \\
				\bottomrule
				
	\end{tabular}}}
\end{table}

In \Cref{table:labeled-datasets-vendors} we show the vendor distribution based
on the labeled dataset (\ie SNMPv3-responsive addresses). To our positive
surprise, more than 82\% of the IPs map to a vendor with a unique
signature.
In total, our dataset covers 16 different vendors. 
We find Cisco to be the dominant router vendor for our labeled dataset with 51\%
of labeled router IPs with unique signatures, followed by Juniper and Huawei
with 10\% each.

For the major router vendors, the majority of IPs can be mapped to unique signatures, which increases our confidence in applying our technique to non-labeled data.
Indeed, this is the case for 
100\% of Juniper, Alcatel/Nokia, and Ericsson router IPs, 
98\% of Cisco router IPs, 
and 86\% of Huawei router IPs. 
Two notable exceptions are MikroTik and H3C. For these two vendors, we might attribute a lower bound of routers. Note that
both these vendors utilize UNIX-based solutions. We use the union of 
signatures in the following sections for router fingerprinting: network 
homogeneity, and end-to-end path analyses.

 \section{ETHICAL CONSIDERATIONS}\label{sec:ethics}

During the design and the application of our methodology we took care to
minimize any potential harm to the operation of routers and networks. First, 
the
load of our measurements is very low. More specifically, we send ten packets,
i.e., one SNMPv3 request and nine probes, three for each one
of ICMP, TCP, and UDP. We do not send any malformed packet to avoid any 
unexpected behavior. 
Moreover, we coordinated with our local network administrators to ensure that
our scanning efforts do not harm the local or upstream network.
We follow current best
practices~\cite{ZMap,partridge2016ethical,dittrich2012menlo} for active 
measurements and ensure that 
our prober IP address has a meaningful DNS PTR record. Additionally, we show
information about our measurements and opt-out possibilities on a website of 
our scanning servers. During our active experiments, we did not receive any
complaints or opt-out requests. Our work uncovers potentially sensitive
security, robustness, and business information about network providers, e.g.,
router vendor. For this we plan to respond to any request by operators 
regarding their networks.
 \section{Router Vendors on a Path}\label{sec:paths}

In this section, we apply \tool to study the diversity of vendors
along data-plane forwarding paths. Such insights are helpful as they could inform routing policy decisions by taking the
equipment on a path into account. For example, if policy or law 
restricts a specific vendor,
\eg \cite{Huawei-is-PRC-spy-ops}, a different path without this vendor
might be selected.
For this analysis, we use the most recent RIPE dataset, namely RIPE-5 (see
Table~\ref{table:ipid-tests-overview}), consisting of 7.3M
traceroutes.

\begin{figure}[!tpb]
    \centering
    \includegraphics[width=.8\linewidth]{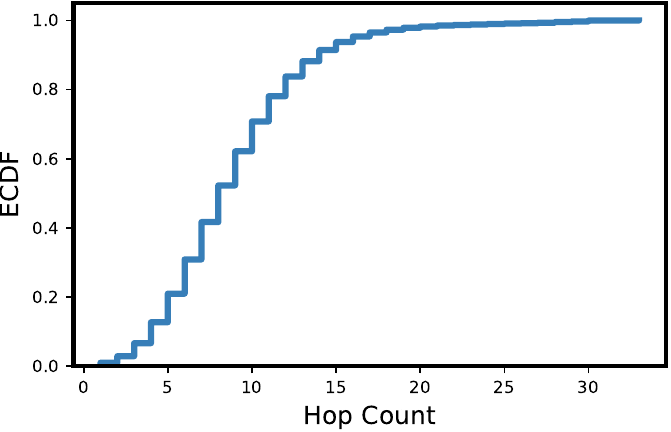}
\vspace{-1em}
    \caption{Path length distribution in the RIPE-5 traceroute dataset.}
\label{fig:ripe5-hop-counts}
\vspace{-1em}
\end{figure}

\begin{figure}[!tpb]
    \centering
    \includegraphics[width=.8\linewidth]{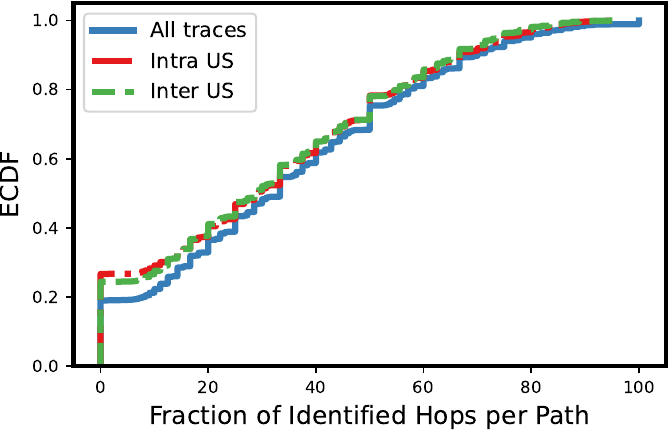}
\vspace{-1em}
    \caption{Identifiable routers on a path (RIPE-5).}
    \label{sec:ripe5-identifiable-routers}
\vspace{-1em}
\end{figure}

\begin{figure}[!tpb]
    \centering
    \includegraphics[width=.8\linewidth]{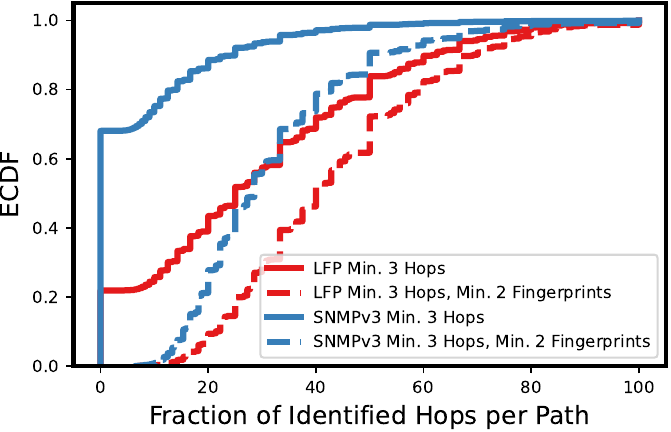}
    \caption{LFP vs. SNMPv3: Identifiable routers distribution on a path.} \label{sec:ripe4-identifiable-routers} 
\end{figure}

Figure~\ref{fig:ripe5-hop-counts} shows the ECDF
of the number of hops per path in the RIPE-5 dataset. In this traceroute
dataset, more than $\approx$ 7.1M of the paths (95\%)
have at least three IP hops. For our analysis, we consider only routable IPv4
addresses and we exclude any addresses that are private, or reserved.  Moreover, more than 95\% of the paths have a length of at most 15 hops.

Figure~\ref{sec:ripe5-identifiable-routers} shows the
fraction of router IPs that we can map to a router
vendor. We notice that for traceroutes with at least three hops, LFP
can identify at least two of the hops in 62\% of the cases.
This fraction increases to 82\% to identify the vendor of at least one hop.
This is a substantial improvement 
compared to the baseline with the SNMPv3 remote router vendor
fingerprinting technique alone, as shown in  Figure~\ref{sec:ripe4-identifiable-routers},
where at least one vendor can be identified for only 35\% of the traceroutes.

\subsection{Identifying Router Vendors on a Path}

\begin{figure}[!tpb]
	\centering
	\includegraphics[width=.8\linewidth]{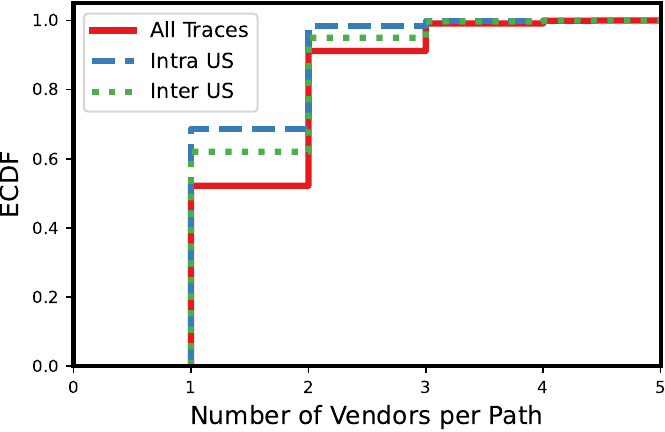}
	\vspace{-1em}
	\caption{Router vendor diversity on a path.}
	\label{fig:ripe5-count-vendors-on-path}
	\vspace{-1em}
\end{figure}

First, we investigate the diversity of router vendors per path as
fingerprinted with \tool.
Figure~\ref{fig:ripe5-count-vendors-on-path} shows the 
number of unique vendors identified on paths where we can identify at least one hop; we 
identify around 650
unique sets of vendors. However, for around 50\% of paths, \tool identifies only a single vendor.
For around 40\% paths, \tool identifies two vendors, 
and only 7\% of the paths have three distinct vendors. 
Four or more different router vendors are identified in fewer than 2\% of the paths.

\begin{figure}[!tpb]
	\centering
	\includegraphics[width=.8\linewidth]{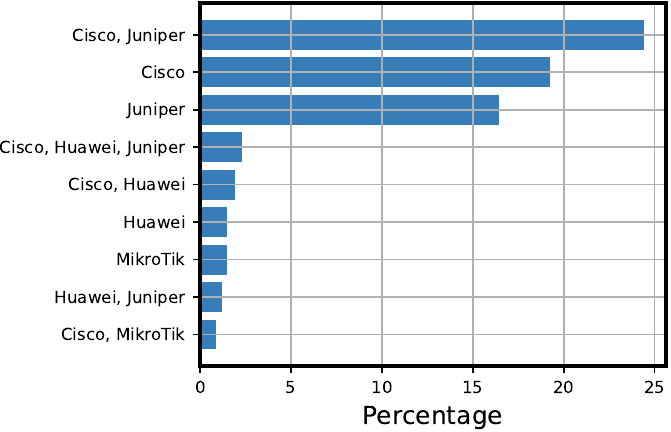}
\vspace{-1em}
	\caption{Top router vendor combinations for paths in the RIPE-5 dataset.}
	\label{fig:tuples-all}
\vspace{-1em}
\end{figure}

Next, we analyze the most popular combinations of router vendors on 
paths (without respect to their order along the path). 
Figure~\ref{fig:tuples-all} shows that
the top nine sets of vendors cover more than 95\% of the RIPE-5 paths.
The top three vendor combinations only involve Cisco and Juniper, making up almost 60\% paths.
Traceroute paths with all other combinations account for fewer than 3\% each.

\subsection{Case Study: US-related Paths}
As a case study, we consider router vendor diversity specifically for the United States. There are 
ongoing discussions whether traffic that originates from the US, or has as a
destination in the US, should be carried by ``untrusted
vendors''~\cite{reuters-untrusted-vendors}. Moreover, if a vulnerability for a
specific router vendor is discovered \cite{CVE-routers, juniper-bugs-cisa, 
cisco_threat_actors},
paths with these vendors might, in theory, be avoided
until a patch is developed and applied. With knowledge about vendors on a forwarding path, possible alternative paths from a source to a destination may receive preferential treatment in routing decisions by network operators.
This could be facilitated with source routing techniques~\cite{IETF-RFC7855}
or enforced by the upstream provider~\cite{Inferring-Complex-AS-Relationships:IMC2014}.

\subsubsection{Intra-US Paths}

\begin{figure}[!tpb]
	\centering
	\includegraphics[width=.8\linewidth]{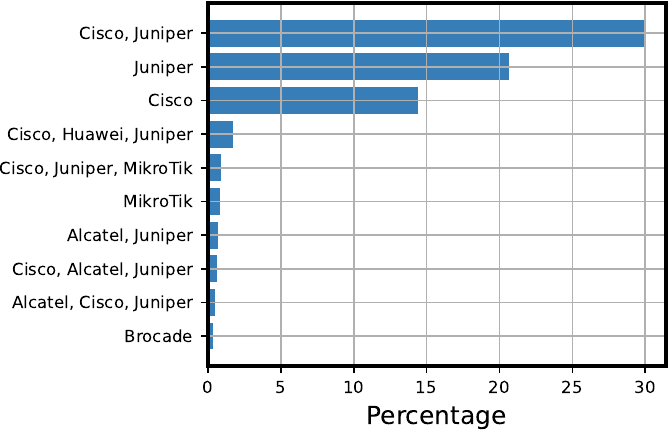}
\vspace{-1em}
	\caption{Top router vendor combinations for intra-US paths in the RIPE-5 dataset.}
	\label{fig:tuples-Intra-US}
\vspace{-1em}
\end{figure}

First, we investigate the case that both the source and the destination of a traceroute are within
the US. To geolocate the endpoints, we rely on IP address registry information.
While other (more fine-grained and more accurate) geolocation techniques exist, 
we are primarily
interested in policies and regulations that are frequently governed by
the home country of the service provider, which is best reflected in
the address registry. 
We exclude from our analysis
anycast IPs~\cite{anycast-IPs} as they may be announced from different 
locations.
The RIPE-5 dataset contains 395,775 traceroutes
with at least three hops where both the source and the destination IP
geolocate to the US. For around 60\%, of them, we can identify two or
more router IPs and assign them to vendors using \tool.
Furthermore, we find that for more than half of intra-US traces we can identify
at least a third of the router vendors on the path.

Moreover, in Figure~\ref{fig:ripe5-count-vendors-on-path}, we show that in around 70\% of the
intra-US paths, all the IPs belong to a single vendor. The majority of the remaining 30\% of intra-US paths have 
routers that belong to two distinct vendors, and the cases where there are three
or more vendors is negligible. Indeed, our results suggest a high degree of 
consolidation of router vendors. Regarding the most popular set of router 
vendors for intra-US paths (cf. \Cref{fig:tuples-Intra-US}), we see a similar 
picture compared to the overall dataset.
Combinations of Cisco and Juniper dominate, even more so than in the
overall dataset, making up more than two thirds of all intra-US paths combined.
This shows that intra-US paths have low vendor diversity, consisting
mostly of Cisco, Juniper, or a combination of both. Such homogeneity
may be indicative of potential critical infrastructure weaknesses
e.g., where all devices are affected by a vulnerability.

\subsubsection{Inter-US Paths}

We also investigate the case that only one of the source and the destination are in
the US. In the RIPE-5 dataset, there are 3M traceroutes
of least three hops where only the source or only the destination IP
geolocate to the US. For around 58\% of these, we can identify the
vendors for two or
more router IPs using \tool.
For more than half of inter-US traces, we can identify the vendor of at least a 
third of the router IPs on the path, 
showing a similar distribution as intra-US as well as other paths.

\begin{figure}[!tpb]
	\centering
	\includegraphics[width=.8\linewidth]{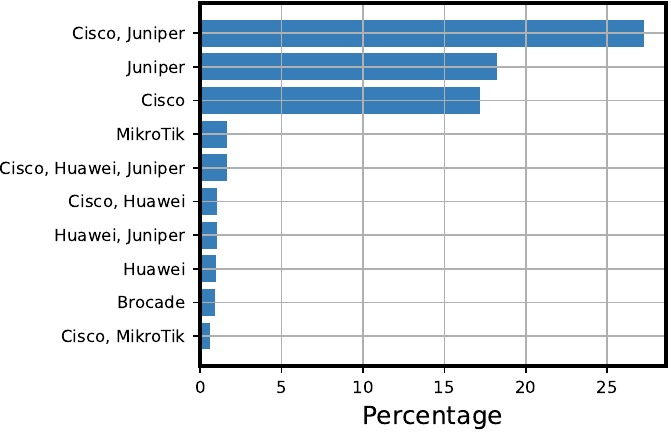}
\vspace{-1em}
	\caption{Top router vendor combinations on inter-US paths (source or destination US).}
	\label{fig:tuples-Inter-US}
\vspace{-1em}
\end{figure}

\begin{table*}[t]
\caption{Two sample unique signatures: top for Juniper and bottom for Cisco. By changing
the default value of Juniper for ICMP iTTL from 64 to 255 (values in box),
the classifier misidentifies Juniper routers as Cisco.}
\label{figure:juniper-cisco}
\vspace{-1em}
\begin{tabular}{l|lllllllllllllll}
Juniper & False & r & r & r & False & False & False & False & 255 &
\fbox{64} & 64 & 84 & 40 &56 & 0\\
Cisco   & False & r & r & r & False & False & False & False & 255 &
\fbox{255} & 64 & 84 & 40 & 56 & 0 \\
\end{tabular}
\end{table*}

Moreover, in Figure~\ref{fig:ripe5-count-vendors-on-path}, we show that in
around 60\% of inter-US paths, all IPs belong to a single vendor. Almost all of the remaining paths contain two mappable router vendors.
These observations are similar to the intra-US study, and show 
a high degree of
vendor consolidation. 
Cisco and Juniper are again the most prominent vendor combinations (see \Cref{fig:tuples-Inter-US}), showing a similar distribution to intra-US and overall paths.
However, the results suggest that inter-US paths
exhibit more 
heterogeneity than the intra-US paths.

\subsection{Case Study: Informed Routing}

Knowing the vendors across the path can inform routing policy. For example, a
sender may want to avoid sending traffic through ASes dominated by
hardware from vendors they do not trust. Thus, the routing policy
could choose an alternative path
if available. Our methodology can inform the possible alternatives and
may serve as a 
step toward enforcing such policies. As a case study, we find
vendor homogeneous ASes
in the RIPE-5 dataset: ASes with at least 1k router IPs where \tool 
finds at least 85\%
of the IPs belong to a single vendor.
Next, we 
use the CAIDA AS relationship dataset~\cite{CAIDA-transit} to find
AS paths where these vendor homogeneous ASes serve as transit ASes. Then, we consider the
destinations ASes where the homogeneous transit AS appears on the
path. For these destinations, we investigate if there exists an alternative 
path
from the \emph{same destination} but with a transit AS using a
\emph{different} vendor. Note that while our analysis utilizes the CAIDA AS
relationships in order to identify policy-compliant transit ASes, such
inferences may be limited by 
the available data and the visibility of \emph{all} AS paths toward a given 
AS. We acknowledge that there may exist paths that cannot be observed from 
publicly available data, or that an alternative path may not be
compliant in the traditional economic or valley-free routing sense.

As a demonstration of the insights possible from this analysis, we
examine two networks: AS9808 and AS3786.  AS9808 is 
a large transit provider where \tool infers Huawei to be the 
the dominant router vendor.  We identify 25,134 AS paths where AS9808 serves as a transit provider.
For 68 destination ASes, no alternative path that does not transit
AS9808 is visible\footnote{Note that not all AS paths are visible in
the BGP
\cite{sigcomm12_ixp,chang2002towards,gregori2014novel,oliveira2009completeness,wohlfart2018leveraging},
thus our analysis is limited to the visible paths only.}. On the other
hand, for 167 destination ASes, an alternative path via
ASes that operate non-Huawei routers is available. 

As a second example for a different router vendor, 
\tool shows that 
Juniper is the dominant router vendor for AS3786. We identify 1.3M AS paths, and 436 
unique destinations where AS3786 appears as a transit provider. For 53 
destinations there is no alternative path visible to us. 
Naturally, our inferences
depend on our visibility into the AS, however, this result suggests
that our methodology can
similarly be applied to any
destination when the set of paths is available.

 \section{Router Fingerprinting}\label{sec:analysis}

With the signatures collected in our active experiments, we now apply our
fingerprinting technique to the router datasets.  We leverage 89 unique
signatures and 78 partial unique signatures from the union dataset (cf.
\Cref{table:ipid-tests-overview,table:labeled-datasets-vendors-protocols}).
Recall that both full and partial unique signatures provide exact matches
between a signature and a vendor.

\begin{figure*}
\begin{minipage}[c]{.32\linewidth}
	\centering
	\includegraphics[width=\linewidth]{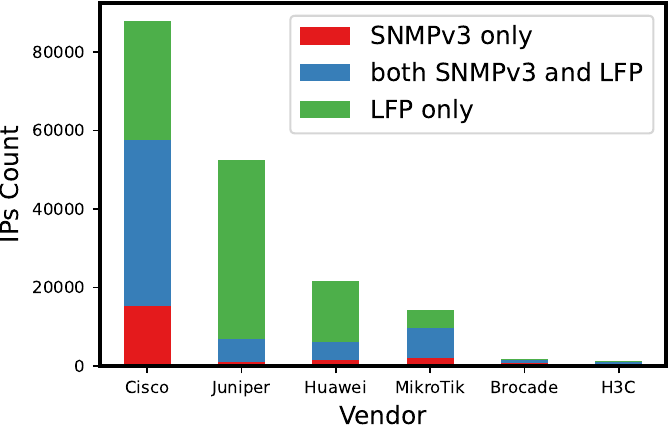}
\vspace{-1em}	
\caption{IPs to vendors: SNMPv3 vs. LFP for the RIPE-5 dataset.}
\vspace{-1em}	
    \label{fig:fp_ip_ripe}
\end{minipage}
\hfill \begin{minipage}[c]{.32\linewidth}
	\centering
	\includegraphics[width=\linewidth]{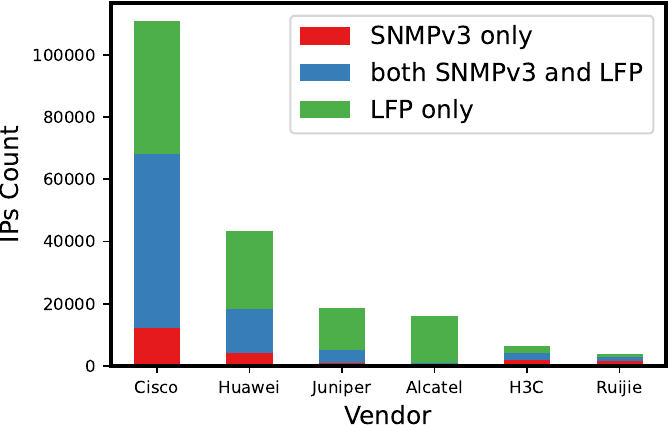}
\vspace{-1em}	
	\caption{IPs to vendors: SNMPv3 vs. LFP for the ITDK dataset.}
\vspace{-1em}	
    \label{fig:fp_ip_itdk}
\end{minipage}
\hfill \begin{minipage}[c]{.32\linewidth}
	\centering
	\includegraphics[width=\linewidth]{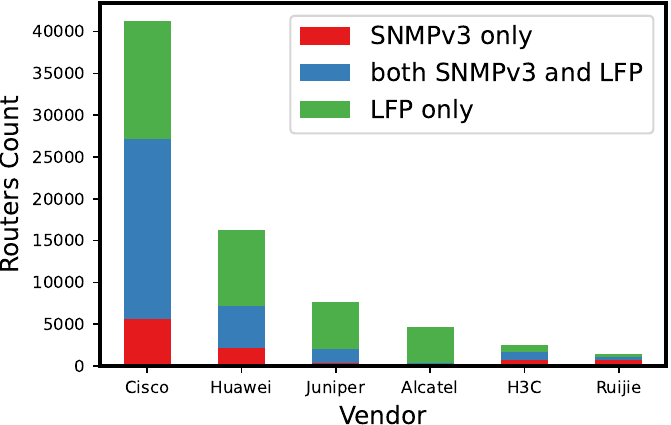}
\vspace{-1em}
	\caption{Routers to vendors: SNMPv3 vs. LFP for the ITDK dataset.}
\vspace{-1em}	
    \label{fig:fp_routers_itdk}
\end{minipage}
\end{figure*}

\subsection{IP to Vendor Mapping}

We use our combined full and partial unique signatures on the latest RIPE
dataset, i.e., RIPE-5, and ITDK
datasets to map IP addresses to vendors.  
For RIPE-5, our analysis shows that our method fingerprints 56.7\% of router IPs when we use unique signatures. For 
reference, the SNMPv3 technique fingerprints only  26\% of the router IPs. LFP alone
fingerprints 49\%. 

\Cref{fig:fp_ip_ripe,fig:fp_ip_itdk} show the fingerprinting results based on
responsive IPs from the RIPE-5 and ITDK datasets, respectively. We report the
router IPs identified only by LFP, only by SNMPv3, and by both methods. We find that
our LFP technique roughly doubles the number of fingerprintable IP addresses for
both datasets.  Moreover, we see that the number of fingerprintable IPs
increases quite drastically for certain vendors: Juniper sees an increase of
650\% and 259.3\%, and Huawei 249.8\% and 136.4\% for RIPE-5 and
ITDK, respectively.  Generally, we see a more balanced router vendor
distribution, with the most dominant vendor Cisco decreasing its share from $\approx$65\%
with SNMPv3 only to $\approx$50\% for SNMPv3 + LFP. We provide an analysis for
the non-unique signature precision and recall in~\Cref{sec:precision-recall}.

\subsection{Router to Vendor Mapping}

Next, we make use of the ITDK dataset not only containing IP address information, but also alias sets.
We apply our signatures to all non-singleton router alias sets.
First, we check if all IPs within fingerprinted alias sets report the same vendor.
We find this to be the case for $\approx 99\%$ of all alias sets, with
498 router IPs producing conflicting vendor inferences (0.65\%).
Second, we plot the router vendor distribution counted by alias set in \Cref{fig:fp_routers_itdk}.
The router distribution is similar to the IP-based distribution (cf. \Cref{fig:fp_ip_itdk}), with Cisco being the dominant vendor, followed by Huawei and Juniper.
Again, we can map about 96.4\% more routers with the combined SNMPv3 + LFP technique, compared to SNMPv3-only.

\subsection{Comparison with other Tools}
\label{sec:analysis:comparison}

To evaluate the accuracy of vendor fingerprinting by LFP and the associated
bandwidth requirements, we conduct a comparison with Nmap \cite{Nmap} and
Hershel~\cite{Hershel}. For this, we acquire a set of addresses from Censys, which are known to reveal vendor information through service banners.
Censys also provides Hershel fingerprints and OS identification where available. 
For each of the top six vendors found via LFP, we randomly select 500 IP addresses and perform tests using both LFP and Nmap.
Additionally, we compare our findings with Hershel fingerprints, wherever 
possible.
\subsubsection{Comparison with Nmap}\label{sec:analysis:comparison:nmap}

\Cref{tab:fpvsnmap} shows the coverage and accuracy results for LFP and Nmap for six different vendors. Coverage refers to the percentage of responsive IPs for each vendor, while accuracy refers to the percentage of correct fingerprints for the responsive IPs. Although both tools have similar accuracy, LFP has the ability to achieve substantially higher coverage.

\begin{table}
	\caption{Comparing coverage and accuracy of LFP and Nmap for Censys-labeled data.}
	\label{tab:fpvsnmap}
	{\begin{tabular}{lrrrrr}
				\toprule
				        & \multicolumn{2}{c}{Coverage} & \multicolumn{2}{c}{Accuracy}  \\
                \cmidrule(lr){2-3}
                \cmidrule(lr){4-5}
				Vendor  & LFP & Nmap & LFP  & Nmap \\
				\midrule
				Cisco   & 40\% &  10\% & 95\%	&  84\%	\\
				Juniper & 81\% &  31\% & 99\%	&  98\%	\\
				Huawei  & 49\% &  20\% & 55\%	&  50\%	\\
				Ericsson& 93\% &  6\% & 80\%	&  0\%	\\
				Mikrotik& 83\% &  15\% & 10\%	&  5\%	\\
				Alcatel	& 38\% &  11\% & 48\%	&  16\%	\\
				
				\bottomrule
	\end{tabular}}
\end{table}

\begin{figure}[t]
	\centering
	\includegraphics[width=.8\linewidth]{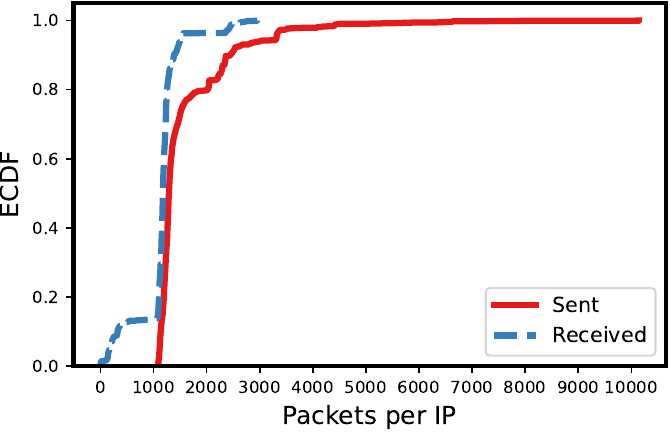}
\vspace{-1em}	
\caption{Sent and received packets from Nmap.}
\vspace{-1em}	
    \label{fig:nmap_packets}
\end{figure}

After evaluating the coverage and accuracy of LFP and Nmap, we
proceeded to analyze their respective bandwidth requirements.
Specifically, LFP sends a consistent 9 packets (3 ICMP, 3 TCP, and 3
UDP) to each targeted IP address. In contrast, Nmap sends substantially more packets when attempting to fingerprint a device. \Cref{fig:nmap_packets} shows the distribution of packets sent and received by Nmap using the default OS fingerprinting options.  
We find that Nmap sends more than 1000 packets to more than 80\% of all targeted IPs.
Moreover, our analysis shows that Nmap sends an average of 1,538 packets per IP and receives 1,065 packets. However, it should be noted that in certain cases Nmap may send an extremely high number of packets to a single IP address, exceeding 10 {\it thousand} packets. This behavior is largely influenced by the services operating on the target system. In contrast, we observe that the LFP technique has considerably lower bandwidth requirements compared to Nmap, making it a more lightweight option overall.

\subsubsection{Comparison with Hershel}
\label{sec:analysis:comparison:hershel}
we also compare LFP with Hershel fingerprints.
By design Hershel requires a single packet to obtain a fingerprint, which is even less bandwidth-intensive than the 9 packets sent by LFP, and much less intrusive than the multitude of packets sent by Nmap.
Our analysis of the test sample shows that Hershel has an overall coverage of approximately 50\%.
Furthermore, we find that Hershel is only able to identify the target vendor with less than 1\% accuracy for our top 3 vendors. This suggests that Hershel---while it may perform well for servers---is not a suitable tool for router fingerprinting.
Additionally, we observe that Hershel often identifies Linux-based systems (such as Mikrotik) simply as Linux machines. This is due to the limited number of signatures for router vendors in the Hershel fingerprinting database.

In summary, LFP achieves a balance between coverage and accuracy while also having a low network footprint.

\subsection{Family-level Fingerprinting}

After discovering that many vendors have not a single but multiple signatures, we investigate whether these signatures can be linked to different router models or families.
To test this hypothesis, we collect a sample of 400 Cisco IPs running
SNMPv2c and query for the Sys.desc O.I.D.~\cite{OID-sys}. This provides a small ground-truth sample with fingerprinting information beyond the vendor. Next, we run LFP against these targets, and collect their signatures. The results show that the collected signatures belong to the top 13 most common Cisco signatures, which cover over 96\% of labeled Cisco data. Additionally, we identify a unique signature for three different IOS families (3 XR, 3 NX, and 7 IOS signatures), which are not shared with the other versions. However, due to the limited ground-truth dataset, it is not possible to evaluate the accuracy of these results in detail, and we leave this task for future work. Overall, the sample data supports the assumption that different signatures can be linked to specific router models or families, which can lead to a more fine-grained router fingerprinting.

\subsection{New Insights on Router Deployment}
\label{sec:analysis:deployment}

Using the collected router fingerprints, we next conduct a comprehensive analysis of global router vendor distribution by comparing our findings with a similar study~\cite{IMC2021-SNMPv3}. Specifically, we utilize \tool to identify the vendor of routers and examine the global distribution of these vendors. Our analysis provides a detailed overview of the global router vendor landscape.

For our analysis, we focus on the ITDK dataset (see
Table~\ref{table:ipid-tests-overview}). Recall that this dataset has information
about all the interfaces (IPs) associated with the same router via alias
resolution. \tool can
identify unique signature routers in 6,743 ASes, compared to 4,929 ASes with the SNMPv3 method.
Thus, not only can LFP identify more than double the number of router IPs (see the previous section), but it
also identifies routers in 1,814 additional ASes (+36.8\%). This is a substantial contribution of
LFP as it sheds light on previous blind spots in the Internet and contributes to a better
estimation of the global router vendor distribution.  

In \Cref{sec:applications} we demonstrate the efficacy of utilizing \tool to enhance router coverage in a network. Our findings reveal that \tool can identify more than twice the number of routers in large networks, thereby substantially improving coverage. Additionally, \tool provides a comprehensive analysis of router homogeneity across different points, offering a more detailed report on homogeneity.

 \section{Discussion}\label{sec:discussion}

\noindent{\bf Obfuscating remote router vendor fingerprinting:} Our analysis shows
that it is possible to hide from remote router fingerprinting. The obvious way
is to drop UDP and TCP traffic, especially from non-whitelisted sources. But
even if UDP and TPC traffic is not dropped, it is still possible to create rare
signatures that are more difficult to be mapped to a specific vendor. It is also
possible by configuring a router to confuse the classification algorithm
(similar to an adversarial attack on classifiers). Some of the features are
difficult to change, e.g., ICMP, TCP, or UDP IPID counters, if they can be configured
at all since they might be directly implemented in the router OS.
However, it is easier to change default iTTL
values. In~\Cref{figure:juniper-cisco}, we present two unique sample
signatures for Juniper (top) and for Cisco (bottom). By changing the default
value of the ICMP iTTL (see Table~\ref{table:features} for details) of the Juniper
routers from 64 to 255, LFP would misclassify the Juniper router as a Cisco router.

\noindent {\bf Using additional sources of information for finger\-pri\-nting:} Our
methodology relies solely on network characteristics and active probing. Other
techniques utilize other sources of information, e.g., banners, that offer good
coverage~\cite{Classifying-Vendors}. Banner data analysis requires the
development of heuristics.
One of the benefits of using a simple rule-based approach such as LFP compared to machine learning (ML) techniques, is that it is clear why certain decisions are being taken, whereas ML techniques usually suffer from a lack of explainability.
Furthermore, complex ML models in networking can suffer from deficits such as shortcut learning, spurious correlations, and vulnerability to out-of-distribution samples \cite{jacobs2022ai}.
Future work should explore the possibility of using explainable ML models for 
router fingerprinting.
Moreover, banners' raw data is less accessible,
typically proprietary, that comes with commercial or limited academic licenses.
Nevertheless, banner data analysis can complement our technique and improve
fingerprinting coverage and granularity. As part of our future work, we plan
to use information fusion of our data and banner data and assess the
benefit of using additional information sources for router fingerprinting, 
especially for vendors with non-unique signature, and
hopefully for finer-grained fingerprinting, e.g., model-level fingerprinting.

\noindent{\bf Non-Unique Signatures:} While we only utilize unique 
signatures in this study, non-unique signatures can offer insights into 
router deployments. This is particularly relevant when a single vendor 
dominates a non-unique signature with thousands of instances. Additionally, 
utilizing non-unique signatures can increase \tool coverage to 64\% 
in the RIPE-5 dataset. We explore the precision and recall of non-unique 
signatures in Appendix~\ref{sec:precision-recall} and intend to investigate additional 
features to enhance the uniqueness of such signatures in future research.

\noindent{\bf Integrating LFP into Nmap:} We also plan to investigate how the
insights gained by our study can be transferred and integrated into
Nmap~\cite{Nmap}. Our analysis shows that LFP can achieve better accuracy with
ten packets (including the SNMPv3 request) than the default Nmap OS detection
mode, which sends up to thousands of probe packets. At least in the case of
router fingerprinting, LFP has proven to be more scalable, less intrusive, and
more accurate. We are already developing a Nmap variation that will replicate
our experiment, and we will share it with the Nmap community to get feedback and
comments. This way, we can improve our methodology and enable more researchers
and engineers to use our technique.

\noindent{\bf Longitudinal analysis:} As part of our future research agenda, we
would like to investigate how we can use our classification methodology and our
collected data to perform a large-scale longitudinal analysis of vendor changes
over time, vendor changes for a network, or vendor changes per router interface
IP. So far, we have collected data that spans more than six months, but the real
potential of our technique will be unveiled by collecting data that spans
multiple years. We plan to publicly make the tools and data available to the
research community and report on our results. We also plan to investigate how
geopolitical events, economic changes, security incidents, and vendor strategies
may influence the distribution of routers by different vendors across different
time scales and geographical regions.

 \section{Conclusion}\label{sec:conclusion} 

In this paper, we have shown that only 10 packet probes per router IP
are enough to accurately fingerprint up to 64\% of routers in the IPv4 Internet.
We developed and evaluated LFP---a lightweight fingerprinting technique that sends three
probe packets for three transport protocols, namely, ICMP, TCP, and UDP. By
augmenting our traces with labeled router data that relies on SNMPv3 responses,
we generated around 90 unique signatures that can accurately identify all major
router vendors. To our surprise, more than half of the routers replied to our
probe packets. The vast majority of the responsive routers (more than 82\%) can
be assigned to only one vendor using our classification. Our results showed that
compared to the state-of-the-art, we more than doubled the coverage of routers
that we can remotely fingerprint, and more accurately inferred the router vendor
compared to popular tools like \nmap. All of this is
achieved with orders of magnitude less probing packets than required by \nmap.
Thus, our mechanism is more scalable, less intrusive, and does not rely on
external and proprietary data like banner grabs. Our classification provides
valuable insights into the deployment of routers within networks and regions,
and the router vendor equipment on a given path. Thus, it can be used to inform routing
decisions, to assesses router deployment strategies, to analyze hardware manufacturer
market share, and to help estimate the potential impact of router vulnerabilities
in a network or a region.  Finally, to enable further research in the area, we
plan to make our LFP tool publicly available.

 \section*{Acknowledgements}\label{sec:ack}

 This work was supported in part by the European
Research Council (ERC) under Starting Grant ResolutioNet (ERC-StG-679158).

\label{page:end_of_main_body}

\balance
\bibliographystyle{ACM-Reference-Format}

\appendix
\section*{Appendix}
\eat{

\section{Ethics}\label{sec:ethics}

During the design and the application of our methodology we took care to
minimize any potential harm to the operation of routers and networks. First, the
load of our measurements is very low. More specifically, we send ten packets,
i.e., one SNMPv3 request and nine transport packet requests, three for each one
of ICMP, TCP, and UDP.
We also randomly distribute our measurements over the address
space, ensuring that at most one packet reaches a target IP each second.
Moreover, we coordinated with our local network administrators to ensure that
our scanning efforts do not harm the local or upstream network.

For the active scanning we use best current
practices~\cite{ZMap,partridge2016ethical,dittrich2012menlo} to ensure that our
prober IP address has a meaningful DNS PTR record. Additionally, we show
information about our measurements and opt-out possibilities on a website of our
scanning servers. During our active experiments, we did not receive any
complaints or opt-out requests.  Our work uncovers potentially sensitive
security, robustness, and business information about network providers, e.g.,
open ports. For this we plan to reach out to network operators and also respond
to requests by operators for their networks.

\eat{
\begin{figure}[!bpt]
	\centering
	\includegraphics[width=\linewidth]{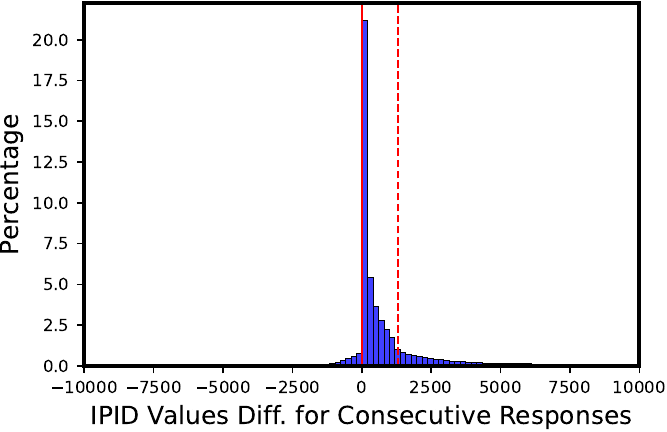}
	\caption{Distribution of IPID difference values for consecutive responses.}
\label{fig:IPID-diff-slide19}
\end{figure}
}

\eat{
\section{IPID Threshold}\label{sec:ipidthreshold}

In~\Cref{fig:IPID-diff-slide19}, we plot the distribution of the IPID difference
values for consecutive responses for the responsive IPs in the RIPE-5 dataset. It
is clear that around 20\% of IPID differences are close to zero.
Close to 90\% of the IPID difference values are included 
by setting a threshold of 1,300, as shown with the dashed vertical line
We use this threshold to differentiate between incremental values and 
random that are dispersed across the full range of possible IPID difference 
values.

Note that an effectively random IPID might by chance fall within this 1,300 threshold.
Since with LFP we take the conservative approach of using the maximum IPID difference between consecutive probes, this random effect is very unlikely to occur twice in a row and thus strongly minimizes the number of false positives.
}

\eat{
\section{Occurrences Threshold}\label{sec:occthreshold}

We perform a sensitivity analysis to understand the impact of the threshold of the
minimum number of occurrences for signatures to be considered.
In~\Cref{fig:o-threshold-slide10}, we vary the threshold (see x-axis),
i.e., the minimum number of IPs with the same signature. A low
threshold leads to a very high number of both unique and non-unique
signatures. This is to be expected as very rare vendors or a small number of
configuration changes may lead to many different signatures.  However, when we
set the threshold of minimum occurrences to consider a signature to 10 or
higher, the number of signatures, both unique and non-unique, converges. As
\Cref{fig:o-threshold-slide10} shows, choosing 10 or 20 as the threshold does not
change the number of signatures substantially. A closer investigation shows that
the set of signatures is also not affected.
Therefore we choose a threshold of 20 router IPs per signature in our study.

}

\eat{
\begin{figure}[!tpb]
	\centering
	\includegraphics[width=\linewidth]{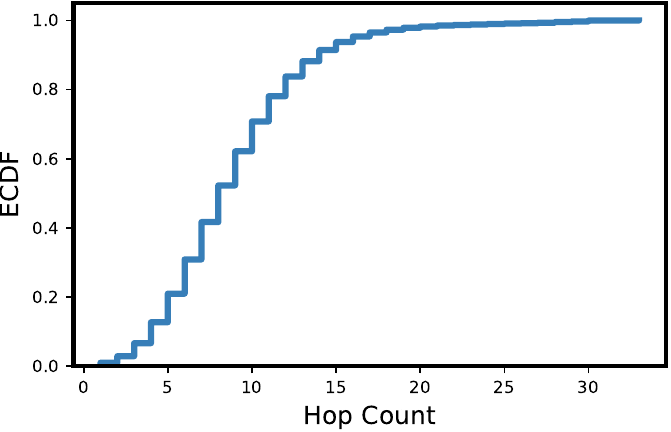}
	\caption{Path length distribution in the RIPE-5 traceroute dataset.}
\label{fig:ripe5-hop-counts}
\end{figure}

\begin{figure}[!bpt]
	\centering
	\includegraphics[width=\linewidth]{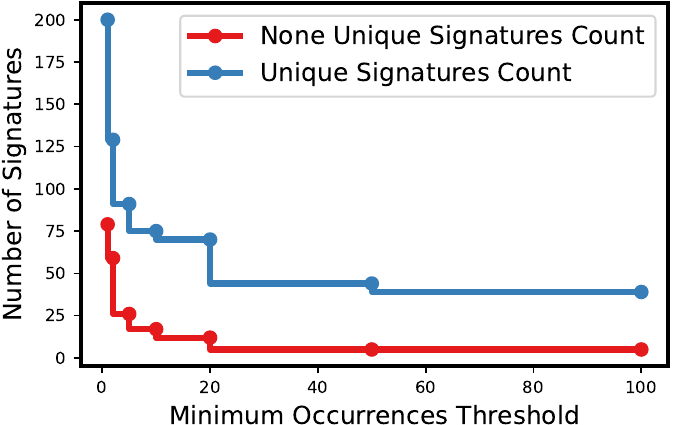}
	\caption{Sensitivity analysis: impact of setting a threshold on the number
of occurrences for signatures on the number of unique and non-unique signatures.}
\label{fig:o-threshold-slide10}
\end{figure}
}

\begin{figure}[!tpb]
	\centering
	\includegraphics[width=\linewidth]{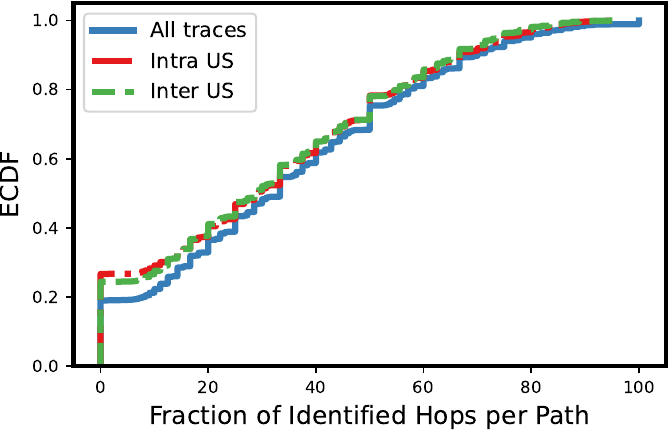}
	\caption{Identifiable routers on a path (RIPE-5).}
	\label{sec:ripe5-identifiable-routers}
\end{figure}

\begin{figure}[!tpb]
	\centering
	\includegraphics[width=\linewidth]{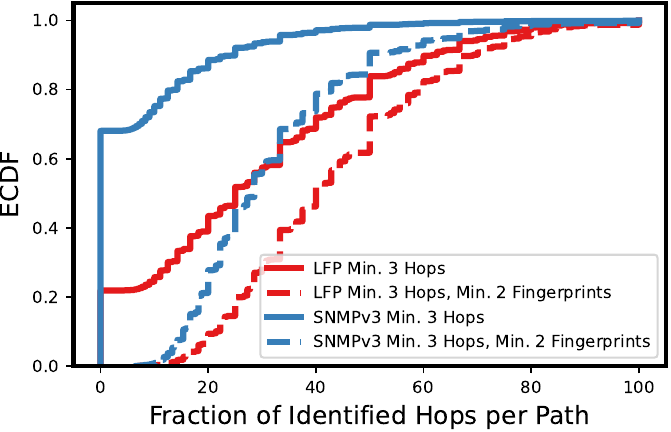}
	\caption{LFP vs. SNMPv3: Identifiable routers distribution on a path.} \label{sec:ripe4-identifiable-routers}
\end{figure}

\section{RIPE-5 Dataset Statistics}\label{sec:appendix-paths}

In Figure~\ref{fig:ripe5-hop-counts} we show the ECDF
of the number of hops per path in the RIPE-5 dataset. In this traceroute
dataset, more than $\approx$ 7.1M of the paths (95\%)
have at least three IP hops. For our analysis, we consider only routable IPv4
addresses and we exclude any addresses that are private, or reserved.  Moreover, more than 95\% of the paths have a length of at most 15 hops.

Figure~\ref{sec:ripe5-identifiable-routers} shows the
fraction of router IPs that we can map to a router
vendor. We notice that for traceroute with at least three hops, LFP can identify at least two hops in 62\% of the cases.
This fraction increases to 82\% to identify the vendor of at least one hop.
This is a substantial improvement 
compared to the baseline with SNMPv3 remote router vendor fingerprinting technique, as shown in  Figure~\ref{sec:ripe4-identifiable-routers},
where at least one vendor can be identified for only 35\% of the traceroutes.
}

\section{Router Vendor Distribution: Revisited}\label{sec:applications}

\begin{figure}[H]
	\centering
	\includegraphics[width=.8\linewidth]{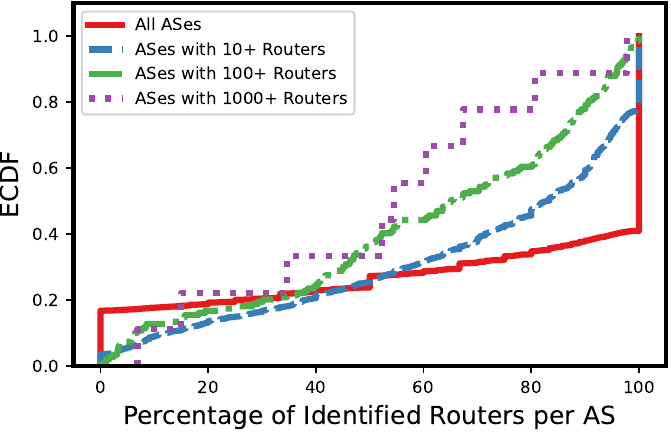}
	\caption{LFP coverage distribution per AS for different minimum 
		thresholds of
		routers per AS.}
	\label{fig:percentage-throshold-routers}
\end{figure}

In Figure~\ref{fig:percentage-throshold-routers} we plot the ECDF of the percentage of
identified routers per network (AS) using LFP. When we consider all the ASes, we find that for
approximately 60\% of the ASes, LFP identifies all the routers. In these ASes we notice a bias: for about half of the ASes there is only one router in the dataset.
When we increase the minimum threshold of routers per AS to consider them in our
study, we notice that for at least 75\% of the ASes LFP identifies at least half
of the routers in an AS. The coverage decreases for large networks, i.e., with
more than 1,000 routers, which is expected as they may have more routers with
closed ports or behind firewalls and other provisions.

\begin{figure}[H]
	\centering
	\includegraphics[width=.8\linewidth]{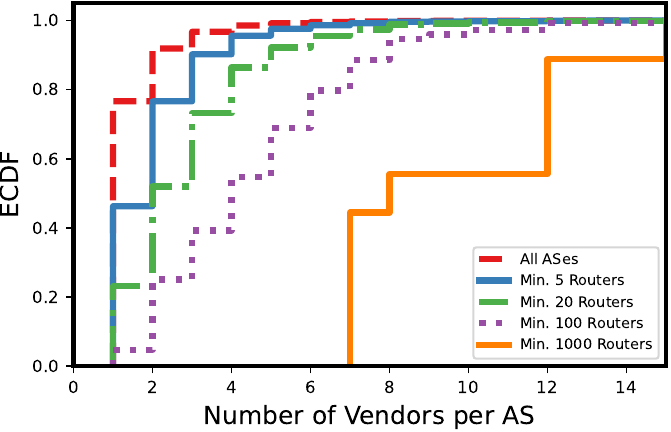}
	\caption{Assessing homogeneity of router vendors per AS.}
	\label{fig:homogeneity-LFP}
\end{figure}

\subsection{Revisiting Homogeneity}

Next, we revisit router vendor homogeneity per AS using LFP. In
Figure~\ref{fig:homogeneity-LFP} we plot the ECDF of the number of
vendors per network (AS). A network is homogeneous when all the routers it hosts
are from the same vendor. Our analysis shows that, indeed, when considering
networks with five routers or more, around half of them have routers of one
vendor and around 75\% with up to two vendors. When we consider larger networks
with more than 20 or 100 routers, we notice that there is only a vendor for
about half and a quarter of networks, respectively. For  large networks, i.e.,
more than 1,000 routers, LFP typically identifies multiple vendors. This is to
be expected as large networks offer multiple services that may require
specialized routers from various vendors. Even a few routers from different
vendors can contribute to the heterogeneity in terms of router vendor per AS.

\subsection{Regional Characteristics}

\begin{figure}[H]
	\centering
	\includegraphics[width=.8\linewidth]{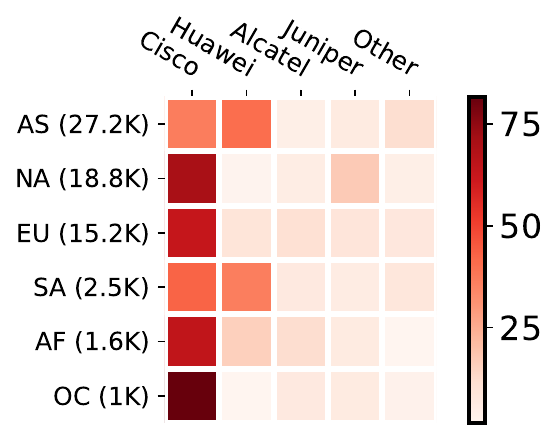}
	\caption{Router vendor popularity per continent.}
	\label{fig:vendors-continent}
\end{figure}

We then study regional characteristics of deployment of router vendors and their
global market share. In Figure~\ref{fig:vendors-continent} we report the number
of routers we can identify with LFP per continent and vendor. The router is
assigned to a location based on the headquarters location of the host network.
Our analysis shows that with LFP, we can double the routers that are identified
in all continents. Overall, the market is very consolidated. A small number of
router manufacturers are responsible for more than 95\% of the routers in a
continent. We notice that in western regions like Oceania, North America (NA),
and Europe, the penetration of Cisco is very high, with 81.7\%,
70.3\%, and 63.2\%, respectively. Cisco also has around 64\% of the market share
in Africa.  Huawei has a substantial market share in Asia and South America,
with 40.6\% and 36.3\%, respectively. Juniper has a  significant market share in
North America, more than 17\%.

In Europe and Asia, the additional contribution of LFP when compared with the
SNMPv3-based fingerprinting is 100\%, i.e., half of the routers could not be
identified with the SNMPv3-based fingerprinting. In North America (NA) and South
America, the additional contribution of LFP is around 87\% and 76\%,
respectively. In North America, one of the reasons is that many Cisco routers
can already be identified with the SNMPv3-based technique and routers in North
America is predominantly Cisco. For South America, the reason is that Huawei
already has a strong presence there, and it can be identified with SNMPv3. The
highest additional contribution of LFP when compared with the SNMPv3-based
fingerprinting is in the two regions with the lower number of identified
routers, namely, Oceania and Africa, with 205\% and 141\%, respectively.

\begin{figure}[h]
	\centering
	\includegraphics[width=.8\linewidth]{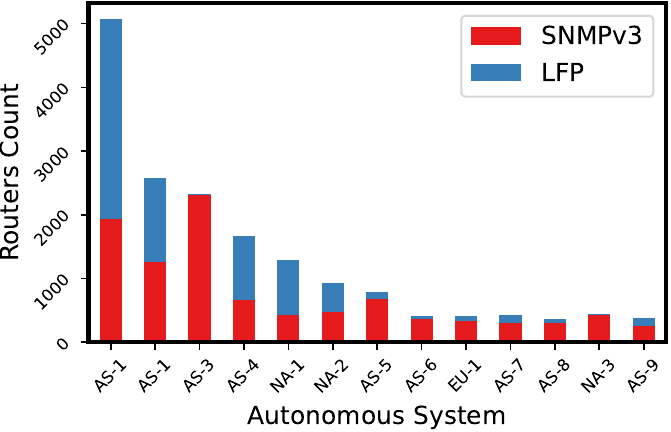}
	\caption{Additional contribution of LFP on router vendor identification in 
		large networks.}
	\label{fig:top-networks}
\end{figure}

Finally, when we turn our attention to the top-13 networks in terms of the
number of routers that we can identify with LFP, we notice that they are spread
around the globe, see Figure~\ref{fig:top-networks}. We also notice that the
additional contribution in identifying routers with LFP compared to SNMPv3-based
fingerprinting varies across the different networks.  Indeed, for the top
network in Asia, the increase is more than 100\%, but for others, e.g., the
third one also in Asia, the additional contribution of LFP is almost
negligible.

\eat{

\begin{table*}[t]
\caption{Brocade: number of instances per signature | unique signature string.}
\label{table:Brocade}
\begin{tabular}{l|lllllllllllllll}
96 & False & i & r & i & False & True & False & False & 64 & 64 & 64 & 84 & 40 &
68 & 0\\
86 & False&i&r&r&False&True&False&False&64&64&64&84&40&68&0\\
44 & False&i&i&r&False&True&False&False&64&64&64&84&40&68&0\\
\end{tabular}
\end{table*}

\begin{table*}[t]
\caption{Cisco: number of instances per signature | unique signature string |
Cisco family.}
\label{table:Cisco}
\begin{tabular}{l|lllllllllllllll|l}
60,747
&True&echo&r&i&False&False&False&False&255&255&255&84&40&56&1&IOS\\
5,387
&True&echo&i&i&False&False&False&False&255&255&255&84&40&96&0&IOS-XR\\
4,976
&True&echo&i&r&False&False&False&False&255&255&64&84&40&56&0&NX-OS\\
3,981
&True&echo&r&i&False&False&False&False&255&255&255&84&40&96&0&IOS-XR\\
1,580
&True&echo&r&i&False&False&False&False&255&255&255&84&40&56&0&IOS\\
1,280
&True&echo&r&r&False&False&False&False&255&255&255&84&40&56&1&IOS\\
924
&True&echo&r&r&False&False&False&False&255&255&255&84&40&96&0\\
598
&False&r&i&r&False&False&False&False&255&255&64&84&40&56&0&NX-OS\\
596
&True&echo&i&i1&False&False&False&False&255&255&255&84&40&96&0\\
444
&True&echo&i&r&False&False&False&False&255&255&255&84&40&96&0\\
317
&True&echo&r&r&False&False&False&False&255&255&64&84&40&56&0&NX-OS\\
275
&True&echo&r&dup&False&False&False&False&255&255&255&40&40&56&0\\
195
&True&echo&i1&i&False&False&False&False&255&255&255&84&40&96&0\\
154
&True&echo&i1&r&False&False&False&False&255&255&64&84&40&56&0\\
112
&True&echo&r&i1&False&False&False&False&255&255&255&84&40&96&0\\
91
&False&sz&r&i&False&False&False&False&255&255&255&56&40&56&1\\
82
&False&i&i&r&False&True&False&False&255&255&64&84&40&56&0\\
52
&False&r&r&dup&False&False&False&False&255&255&255&40&40&56&0\\
49
&True&echo&sz&r&False&False&False&False&255&255&64&84&40&56&0\\
49
&True&echo&r&dup&False&False&False&False&255&255&255&40&40&96&0\\
37
&True&echo&r&i&False&False&False&False&255&255&255&84&40&168&1\\
30
&True&echo&i&i&False&False&False&False&255&255&255&84&40&96&1\\
24
&False&r&r&r&False&False&False&False&255&255&64&84&40&56&0\\
20
&True&echo&r&i&False&False&False&False&255&255&255&84&40&96&1\\
20
&True&echo&r&dup&False&False&False&False&255&64&64&40&40&56&0\\
\end{tabular}
\end{table*}

\begin{table*}[t]
\caption{Ericsson: number of instances per signature | unique signature string.}
\label{table:Ericsson}
\begin{tabular}{l|lllllllllllllll}
200&True&echo&i&i&False&False&False&True&255&255&64&84&40&56&0
\end{tabular}
\end{table*}

\begin{table*}[t]
\caption{H3C: number of instances per signature | unique signature string.}
\label{table:H3C}
\begin{tabular}{l|lllllllllllllll}
276 &
False&i&r&r&False&True&False&False&255&255&255&84&40&56&0\\
37 &
False&i&i&r&False&True&False&False&255&255&255&84&40&56&0\\
23 &
False&r&r&r&False&False&False&False&255&255&255&84&40&56&0\\
22 &
False&i&r&r&False&True&False&False&255&255&255&84&40&68&0\\
\end{tabular}
\end{table*}

\begin{table*}[t]
\caption{Huawei: number of instances per signature | unique signature string.}
\label{table:Huawei}
\begin{tabular}{l|lllllllllllllll}
14,827 &
True&echo&i&i&False&False&False&True&255&255&255&84&40&56&0\\
1,208 &
True&echo&i&i&False&False&False&False&255&255&255&84&40&56&0\\
460 &
True&echo&r&r&False&False&False&False&255&255&255&84&40&56&0\\
303 &
True&echo&i&i1&False&False&False&False&255&255&255&84&40&56&0\\
129 &
True&echo&i1&i&False&False&False&False&255&255&255&84&40&56&0\\
60 &
True&echo&i&dup&False&False&False&False&255&255&255&40&40&56&0\\
24 &
True&echo&r&i1&False&False&False&False&255&255&255&84&40&56&0\\
23 &
False&r&i&i&False&False&False&True&255&255&255&84&40&56&0\\
\end{tabular}
\end{table*}

\begin{table*}[t]
\caption{Juniper: number of instances per signature | unique signature string.}
\label{table:Juniper}
\begin{tabular}{l|lllllllllllllll}
6,425 &
False&i&i&i&True&True&True&True&255&64&64&84&40&56&0\\
4,260&
False&r&r&r&False&False&False&False&255&64&64&84&40&56&0\\
3,919&
False&i&i&i&False&True&False&True&255&64&64&84&40&56&0\\
1,531&
False&r&i&i&False&False&False&True&255&64&64&84&40&56&0\\
480&
False&r&r&i&False&False&False&False&255&64&64&84&40&56&0\\
296&
False&i&r&i&False&True&True&False&255&64&64&84&40&56&0\\
271&
True&echo&i&i&False&False&False&True&255&64&64&84&40&56&0\\
130&
True&echo&i&sz&False&False&False&False&128&128&64&84&40&56&0\\
89&
False&i&r&r&False&True&False&False&255&64&64&84&40&56&0\\
75&
False&i&i&r&False&True&False&False&255&64&64&84&40&56&0\\
56&
False&sz&sz&sz&False&False&False&False&64&64&64&84&40&68&0\\
42&
False&r&i&r&False&False&False&False&255&64&64&84&40&56&0\\
42&
False&i&i&i&True&True&True&True&255&64&64&84&40&56&1\\
29&
False&r&r&r&False&False&False&False&255&64&64&84&40&56&1\\
20&
True&echo&r&sz&False&False&False&False&128&128&64&84&40&56&0\\
\end{tabular}
\end{table*}

\begin{table*}[t]
\caption{MikroTik: number of instances per signature | unique signature string.}
\label{table:Mikrotik}
\begin{tabular}{l|lllllllllllllll}
1,941 &
False&i&sz&dup&False&True&False&False&64&64&64&84&40&68&0\\
1,648&
False&dup&sz&dup&False&False&False&False&64&64&64&84&40&68&0\\
1,427&
False&snz&sz&dup&False&False&False&False&64&64&64&84&40&68&0\\
972&
False&snz&sz&snz&False&False&False&False&64&64&64&84&40&68&0\\
649&
False&i1&sz&dup&False&True&False&False&64&64&64&84&40&68&0\\
618&
False&dup&sz&snz&False&False&False&False&64&64&64&84&40&68&0\\
488&
False&r&sz&dup&False&False&False&False&64&64&64&84&40&68&0\\
250&
False&r&r&dup&False&False&False&False&64&64&64&40&58&68&0\\
225&
False&r&dup&r&False&False&False&False&64&64&64&40&40&68&0\\
221&
False&i&sz&dup&False&True&True&False&64&64&64&84&40&68&0\\
206&
False&i1&sz&snz&False&True&False&False&64&64&64&84&40&68&0\\
192&
False&dup&sz&r&False&False&False&False&64&64&64&84&40&68&0\\
80&
False&dup&sz&i&False&False&False&False&64&64&64&84&40&68&0\\
58&
False&r&dup&dup&False&False&False&False&64&64&64&40&40&68&0\\
56&
False&i&sz&i&False&True&False&False&64&64&64&84&40&68&1\\
53&
True&echo&dup&r&False&False&False&False&64&64&64&40&40&68&0\\
53&
False&i1&sz&dup&False&True&True&False&64&64&64&84&40&68&0\\
46&
False&r&r&dup&False&False&False&False&64&64&255&40&56&68&0\\
43&
False&r&dup&dup&False&False&False&False&64&64&64&40&58&68&0\\
42&
False&r&r&r&False&False&False&False&64&64&64&40&40&68&0\\
40&
False&r&r&r&False&False&False&False&255&255&64&40&58&56&0\\
39&
False&r&dup&dup&False&False&False&False&64&64&64&58&40&84&0\\
27&
False&i&r&r&False&True&False&False&255&255&64&40&58&56&0\\
26&
False&i&r&r&False&True&False&False&255&255&255&40&56&56&0\\
25&
False&r&dup&i&False&False&False&False&64&64&64&40&40&68&0\\
22&
False&r&r&r&False&False&False&False&255&255&255&40&56&56&0\\
22&
False&i&r&dup&False&True&False&False&64&64&64&40&58&68&0\\
20&
False&snz&sz&r&False&False&False&False&64&64&64&84&40&68&0\\
\end{tabular}
\end{table*}

\begin{table*}[t]
\caption{net-snmp: number of instances per signature | unique signature string.}
\label{table:net-snmp}
\begin{tabular}{l|lllllllllllllll}
194&
False&r&sz&i&False&False&False&False&64&64&64&84&40&68&0\\
31&
False&sz&sz&i&False&False&False&False&64&64&64&84&40&68&0\\
\end{tabular}
\end{table*}

\begin{table*}[t]
\caption{Ruijie: number of instances per signature | unique signature string.}
\label{table:Ruijie}
\begin{tabular}{l|lllllllllllllll}
46&False&i&r&r&False&True&False&False&64&64&64&84&56&56&0\\
\end{tabular}
\end{table*}

}

\section{Precision and Recall}\label{sec:precision-recall}

For the labeled RIPE-6 data with SNMPv3 information, we perform a
precision and recall study. We do a 80/20 random split where we use 80\% of 
the
data for training and the other 20\% for testing.
The results per
vendor for precision and recall are presented 
in~\Cref{table:Precision-Recall}.
For the major vendors, namely, Cisco, Juniper, and Huawei, both precision and
recall are very high, close to 1. Precision is also high for popular vendors,
e.g.,  Nokia, Ruijiem, and Ericsson, but the recall is lower. We attribute 
this
to the relatively low testing sample. The precision and especially recall is
very low for UNIX-based vendors, e.g., net-snmp, Brocade, and H3C.

\begin{table}[t]
	{\scriptsize
		\center
		\caption{Precision and Recall: data random split (80/20)}
		\label{table:Precision-Recall}
		\resizebox{.8\columnwidth}{!}{
			\begin{tabular}{lrrr}
				\toprule
				Vendor &  Recall & Precision & Total (test)  \\
				\midrule
				Cisco & 0.99 & 0.99 &6,754\\
				Mikrotik & 0.99 & 1.0 &919\\
				Juniper & 0.97 & 0.99 & 789\\
				Huawei & 0.96 & 0.98 & 450\\
				Brocade & 0.64 & 0.72 & 153\\
				H3C  & 0.20 & 0.23 & 123\\
				Nokia & 0.8 & 1 &  64\\
				Ruijie & 0.77 & 1 & 10\\
				Ericsson & 0.77 & 1 & 10\\
				net-snmp & 0.35 & 0.37 & 315\\
				\bottomrule
	\end{tabular}}}
\end{table}

\end{document}